\documentclass[11pt, onecolumn, draftnocls]{IEEEtran}
\IEEEoverridecommandlockouts
\usepackage{amsfonts,booktabs}
\usepackage{amsmath}
\usepackage{graphicx}
\usepackage{subfigure}
\usepackage{times}
\usepackage{geometry}
\usepackage{setspace}
\usepackage{pgfplots}
\usepackage{pgfplotstable}
\usepackage[subnum]{cases}
\usepackage{filecontents}
\usepackage{multirow}
\usepackage{color}
\let\oldbibliography\thebibliography
\renewcommand{\thebibliography}[1]{%
  \oldbibliography{#1}%
  \setlength{\itemsep}{0pt}%
}

\newenvironment{remark}         {\begin{sideremark}\rm}{\end{sideremark}}
\newtheorem{sideremark}{Remark}

\newtheorem{definition}         {Definition}[section]
\newtheorem{lemma}{Lemma}
\newtheorem{theorem}{Theorem}

\def\reals{{\rm I\kern-.17em R}}
\def\nats{{\rm I\kern-.17em N}}

\newcommand{\beq}{\begin{equation}}
\newcommand{\eeq}{\end{equation}}
\newcommand{\beqa}{\begin{eqnarray}}
\newcommand{\eeqa}{\end{eqnarray}}

\newcommand{\mathbi}[1]{\ensuremath \textbf{\em #1}}
\newcommand{\paren}[1]{\left(#1\right)}
\newcommand{\sqparen}[1]{\left[#1\right]}
\newcommand{\brparen}[1]{\left\{#1\right\}}
\newcommand{\field}[1]{\ensuremath{\mathbb{#1}}}
\newcommand{\abs}[1]{\left|#1\right|} 
\newcommand{\N}{\ensuremath{\field{N}}} 
\newcommand{\R}{\ensuremath{\field{R}}} 
\newcommand{\I}[1]{\ensuremath{\mathsf{1}_{\left\{#1\right\}}}} 
\newcommand{\ra}{\ensuremath{\rightarrow}} 
\newcommand{\PR}[1]{\ensuremath{\mathsf{Pr}\left\{#1\right\}}} 
\newcommand{\PRP}[1]{\ensuremath{\mathsf{Pr}\left(#1\right)}} 
\newcommand{\EW}{\ensuremath{\mathsf{E}}} 
\newcommand{\ES}[1]{\ensuremath{\mathsf{E}\left[#1 \right]}} 
\newcommand{\e}[1]{\ensuremath{{\rm e}^{#1}}} 

\newcommand{\BO}[1]{\ensuremath{O\paren{#1}}}
\newcommand{\LO}[1]{\ensuremath{o\paren{#1}}}
\newcommand{\BT}[1]{\ensuremath{\Theta\paren{#1}}}
\renewcommand{\vec}[1]{\ensuremath{\boldsymbol{#1}}}

\newcommand{\logp}[1]{\ensuremath{\log\paren{#1}}}
\newcommand{\inv}[2]{\ensuremath{{#1}^{-1}\paren{#2}}}

\newcommand{\PV}{\mathbi{P}\paren{\mathbi{h},\mathbi{g}}}
\newcommand{\PS}[1]{P_{#1}\paren{\mathbi{h},\mathbi{g}}}
\newcommand{\OPC}[2]{\ensuremath{P_{#1,#2}^\star\paren{ \mathbi{h}, \mathbi{g} }}}
\newcommand{\SPAP}[2]{\ensuremath{{\hat{{\vec{P}}}^{\rm {#1}}_{#2}}}}
\newcommand{\SPC}[2]{\ensuremath{\hat{P}^{\rm {#1}}_{i,{#2}}}}

\newcommand{\X}[2]{\ensuremath{X_N^\star\paren{ #1,  #2}}}
\newcommand{\Xr}[2]{\ensuremath{X_{K_N}^\star\paren{ #1,  #2}}}
\newcommand{\Xrt}[2]{\ensuremath{\tilde{X}_{K_N}^\star\paren{#1, #2}}}
\newcommand{\Xt}[2]{\ensuremath{\tilde{X}^\star_N\paren{ #1,  #2}}}
\newcommand{\h}[1]{\ensuremath{h^\star_{#1}}}
\newcommand{\z}[1]{\ensuremath{z_{#1}}}
\newcommand{\g}[2]{\ensuremath{g_{#1:#2}}}
\newcommand{\K}{\ensuremath{K_{N}}}

\newcommand{\CoSTPILFull}{\text{CoS}^{\rm F}_{\rm TPIL}}
\newcommand{\CoSTPILRed}{\text{CoS}^K_{\rm TPIL}}
\newcommand{\CoSIPILFull}{\text{CoS}^{\rm F}_{\rm IPIL}}
\newcommand{\CoSIPILRed}{\text{CoS}^K_{\rm IPIL}}
\newcommand{\CoSILFull}{\text{CoS}^{\rm F}_{\rm IL}}
\newcommand{\CoSILRed}{\text{CoS}^K_{\rm IL}}
\newcommand{\RateTPILFull}{\ensuremath{R^{\rm F}_{\rm TPIL}\paren{N}}}
\newcommand{\RateTPILRed}{\ensuremath{R^{K}_{\rm TPIL}\paren{K_N}}}
\newcommand{\RateILFull}{\ensuremath{R^{\rm F}_{\rm IL}\paren{N}}}
\newcommand{\RateILRed}{\ensuremath{R^K_{\rm IL}\paren{K_N}}}
\newcommand{\RateIPILFull}{\ensuremath{R^{\rm F}_{\rm IPIL}\paren{N}}}
\newcommand{\RateIPILRed}{\ensuremath{R^K_{\rm IPIL}\paren{K_N}}}
\newcommand{\RateIPM}{\ensuremath{R_{\rm IPL}\paren{N}}}
\newcommand{\RateTPM}{\ensuremath{R_{\rm TPL}\paren{N}}}
\newcommand{\SRate}[2]{\ensuremath{\hat{R}_{\rm #1}\paren{#2}}}
\newcommand{\textbox}[5]{\node at (#1,#2) [rectangle,draw, fill=white,text width=#3, text =black] {\center \tiny \bf STSB-FM: \bf{#4} STPB-FM: \bf{#5}};}
\newcommand{\pfgsetting}{\pgfplotsset{
width= 8cm, 
every axis/.append style={line width=1.2pt},
label style={font=\bf\scriptsize}, 
ylabel style={yshift=-0.8em},
xlabel={Number of Secondary Users},
ylabel={Throughput (nats per channel use)},
title style={font=\bf\scriptsize}, 
tick label style={font=\scriptsize,/pgf/number format/1000 sep={} },
tick style={ line width=1.5pt},
legend style={font=\bf\tiny,cells={anchor=west}},
every mark/.append style={solid}
}}
\newcommand{\axissetting}{ xmin=0,xmax=1000,xtick={0 ,100,200,300,400,500,600,700,800,900,1000},grid=major, tick style={color=black, major tick length={0.10 cm}}, grid style={line width= 0.75pt, densely dotted, color= black}}

\begin{document}

\title{Multiuser diversity for the cognitive uplink with generalized fading and reduced primary's cooperation}

\author{Ehsan Nekouei, \IEEEmembership{Student Member, IEEE}, Hazer Inaltekin, \IEEEmembership{Member, IEEE} and Subhrakanti Dey, \IEEEmembership{Senior Member, IEEE} 
\thanks{E. Nekouei and S. Dey are with the Department of Electrical and Electronic Engineering, The University of Melbourne, VIC 3010, Australia. (e-mails: e.nekouei@pgrad.unimelb.edu.au and sdey@unimelb.edu.au) 

H. Inaltekin is with the Department of Electrical and Electronics Engineering, Antalya International University, Antalya, Turkey. (e-mail: hazeri@antalya.edu.tr).}}
\maketitle
\thispagestyle{empty}
\vspace{-1cm}
\begin{abstract}
In cognitive multiple access networks, feedback is an important mechanism to convey secondary transmitter primary base station (STPB) channel gains from the primary base station (PBS) to the secondary base station (SBS). This paper investigates the optimal sum-rate capacity scaling laws for cognitive multiple access networks in feedback limited communication scenarios. First, an efficient feedback protocol called $K$-smallest channel gains ($K$-SCGs) feedback protocol is proposed in which the PBS feeds back the $\K$ smallest  out of $N$ STPB channel gains to the SBS. Second, the sum-rate performance of the $K$-SCG feedback protocol is studied for three network types when transmission powers of secondary users (SUs) are optimally allocated.  The network types considered are total-power-and-interference-limited (TPIL), interference-limited (IL) and individual-power-and-interference-limited (IPIL) networks.  For each network type studied, we provide a sufficient condition on $\K$ such that the $K$-SCG feedback protocol is {\em asymptotically} optimal in the sense that the secondary network sum-rate scaling behavior under the $K$-SCG feedback protocol is the same with that under the full-feedback protocol. We allow distributions of secondary-transmitter-secondary-base-station (STSB), and STPB channel power gains to belong to a fairly general class of distributions called class $\mathcal{C}$-distributions that includes commonly used fading models.  It is shown that for $\K=N^\delta$ with $\delta\in\left(0,1\right)$, the $K$-SCG feedback protocol is asymptotically optimal in TPIL networks, and the secondary network sum-rate scales according to $\frac{1}{n_h}\log\logp{N}$, where $n_h$ is a parameter obtained from the distribution of STSB channel power gains. In this case, it is also shown that the average interference power at the PBS can be made arbitrarily small without losing anything from this optimal sum-rate scaling behavior. For IL networks, the $K$-SCG feedback protocol is asymptotically optimal if $\K=\BO{1}$. In this case, the secondary network sum-rate scales optimally according to $\frac{1}{\gamma_g}\logp{N}$, where $\gamma_g$ is a parameter obtained from the distribution of STPB channel gains. Finally, for IPIL networks, it is proven that the $K$-SCG feedback protocol is asymptotically optimal if $\K=\BO{1}$, and the secondary network sum-rate scales according to $\min\paren{1,\frac{1}{\gamma_g}}\logp{N}$. An extensive simulation and numerical study is also performed to illustrate the established sum-rate capacity scaling laws for finite networks.    
\end{abstract}

\section{Introduction}
\subsection{Background and Motivation}

The electromagnetic radio spectrum is a limited communication resource. This fact makes its allocation and exploitation one of the prime concerns to accommodate increasingly more data-rate-intense wireless services and next generation wireless systems in today's already vastly crowded spectrum. Part of the reason for the perceived crowdedness of the radio spectrum is the current practice of managing it, which is the legacy {\em command-and-control} regulation \cite{FCC02}. This is a static regulatory approach to the spectrum management based on exclusive usage rights assigned to a number of licensees. Being static by its nature, the command-and-control approach cannot utilize spatio-temporal spectrum usage characteristics of the incumbent users (alternatively called: primary users or PUs), and therefore cannot opportunistically assign radio spectrum to other unlicensed third parties (alternatively called: cognitive users, secondary users or SUs). 

As a response to the sheer pressure of having a more dynamic means of managing spectrum and exploiting likely spectrum holes, there emerged the idea of cognitive radio technology as a revolutionary ``disruptive but unobtrusive"  technique \cite{Mitola99b}. Roughly speaking, it alleviates the spectrum scarcity problem by allowing cognitive users to share the same bandwidth with the PUs provided that their transmissions do not cause harmful degradation to the primary transmission. Various signal processing, information-theoretic and protocol related aspects of the cognitive radio networks have been investigated extensively over the last decade, {\em e.g.,} see \cite{Haykin05}-\cite{Akyildiz06} and references therein.  Among many other issues, one of the recurrent themes appearing in most of the earlier papers on cognitive radio is the awareness of cognitive transmitters (or, the awareness of a central entity to perform scheduling and resource allocation among SUs) about the channel states of PUs and SUs.  

Briefly, channel side information at the cognitive transmitters is necessary for the proper completion of the cognition cycle, and the harmonious operation of PUs and SUs in a given frequency band.  However, for large numbers of SUs, this requirement puts an excessive and impractical feedback burden on the backhaul link between the primary and secondary networks, which leads to the following research question of interest here: What are the optimal capacity scaling laws for cognitive secondary networks containing large numbers of SUs when secondary-transmitter-primary-base-station (STPB) channel states are {\em only partially} available? The current paper addresses this question for the specific case of a cognitive multiple access (CMAC) network in which the sum-rate capacity of the CMAC network is the primary performance figure of interest, and the STPB channel states are partially available at the secondary base-station (SBS). Here, ``partially available" means only a subset of channel states are available at the SBS.  The SBS performs jointly optimum power control and scheduling policy to extract the maximum possible sum-rate from the SUs based on the available channel side information and subject to interference power constraints at the primary base-station (PBS).

More specifically, we consider a CMAC network in which $N$ SUs transmit data to a common SBS, and interfere with the signal reception at a PBS at the same time.  This is the commonly used underlay paradigm for the coexistence of primary and secondary networks \cite{Goldsmith09}, which is also known as the spectrum sharing technique \cite{Ghasemi07}.  For such communication instances, availability of STPB channel gains of {\em all} SUs at the secondary network is a frequent assumption in the cognitive radio literature, {\em e.g.,} see \cite{RZhang09}-\cite{NID12}.  While this assumption is crucial for the SBS to manage the secondary network interference power at the PBS as well as to implement jointly optimum power control and scheduling policy, it places an extra burden on the primary network.  In order STPB channel gains to be available at the SBS, the PBS should estimate STPB channel gains for all SUs and convey them to the SBS by means of a primary-secondary feedback link (PSFL), \emph{i.e.,} backhaul link, or by means of a band manager mediating communication between primary and secondary networks \cite{Ghasemi07, Musavian09,Peha05}.  In either case, the feedback is required, and the capacity of the feedback link is limited in general.  Hence, for large numbers of SUs, it becomes impractical for the PBS to convey all STPB channel gains to the SBS within the channel coherence time due to various physical restrictions on the communication system of interest such as feedback link capacity limitations and energy constraints. 

 
Here, we show that the capacity of the backhaul link does not act as a primary bottleneck on the sum-rate capacity scaling behavior of CMAC networks if the user selection for feedback is performed intelligently at the PBS.  That is, we find that the secondary networks can achieve the optimal sum-rate capacity scaling laws even when the STPB channel gains are partially available at the SBS for some of the strategically chosen SUs, {\em i.e.,} see Theorems \ref{Theo: TPIL}, \ref{Theo: IL} and \ref{Theo: IPIL} in Section \ref{Sec: R&D} for greater details.  Although the total feedback load ({\em i.e.,} the number of  SUs whose channel states to be fed back to the SBS) grows large with the total number of SUs, the rate of increase of the feedback load can be made arbitrarily smaller than the rate of increase of the total number of SUs.  From a practical point of view, this finding implies a significant primary-secondary feedback load reduction for cognitive radio networks with large numbers of SUs without any significant first-order reduction in the secondary network data rate.  Our results further indicate that secondary networks can coexist with primary networks by causing {\em almost} no interference, \emph{i.e.,} see Theorem \ref{Theo: Interference Convergence} in Section \ref{Sec: R&D}.  Through the characterization of the scaling behavior of the secondary network sum-rate in terms of feedback link capacity, fading parameters and the number of SUs, our results shed light on the fundamental tradeoffs between the secondary network sum-rate capacity and the feedback link capacity.  They also provide critical engineering insights for the design of primary-secondary feedback protocols and for cognitive radio network planning. 

In this paper, all of our results are derived for a parametrized family of general fading distributions called class $\mathcal{C}$-distributions. The available tools, in  the literature, for analyzing multiuser diversity gain (MDG) in cognitive radio networks are applicable only if one can find a closed form expression for the joint channel states. This is not always possible when direct and interference channel gains are arbitrarily distributed, \emph{e.g.,} when direct and interference channel gains are Nakagami-$m$ distributed. Dealing with a parametrized family of distributions, to derive generalized MDGs, requires investigation of more subtle concentration behavior of extreme order statistics to obtain tail estimates of joint channel states which is technically much more challenging than assuming specific distributions for direct and interference channel gains.

In addition to being technically challenging, perhaps more importantly, our analysis provides new insights into the network operation by relating the fading distribution parameters to the pre-log factors in the derived throughput scaling laws, which is otherwise hidden by assuming specific fading processes such as Rayleigh fading.

\subsection{Main Contributions in Detail} 
Design of feedback reduction policies for the PSFL is a challenging issue due to dependence of multiuser diversity gains and the interference management task on the knowledge of STPB channel gains. We consider that the PBS is able to send the STPB channel gains of at most $\K $ SUs to the SBS, where $\K$ is an integer smaller than or equal to $N$ and possibly changing as a function of $N$.\footnote{Although we do not show the dependence of $K$ on $N$, it should be understood that $K$ is a function of $N$ in the remainder of the paper.}  $\K=N$ case is named as the {full}-feedback protocol in which the SBS has the perfect knowledge of all STPB channel gains to implement user scheduling and power control.  $\K$ can be interpreted as our modeling parameter to numerically designate the feedback capability of the PSFL, {\em i.e.,} the more capacity the PSFL has, the larger $\K$ is.  In the absence of any knowledge about the secondary-transmitter-secondary-base-station (STSB) channel gains, the best strategy for the PBS is to pick the least harmful SUs by feeding back the channel gain $g_i$ of the SU-$i$ if and only if $g_i\leq \g{\K }{N}$, where $\g{\K }{N}$ is the $\K$th smallest value in the set $\left\{{g_i}\right\}_{i=1}^N$.  Formally speaking, $g_{\K:N}$ is the $\K$th order statistic for the collection of random variables $\left\{g_i\right\}_{i=1}^N$. We refer to this feedback policy as the $K$-\emph{smallest channel gain} ($K$-SCG) feedback protocol. Hence, using the $K$-SCG feedback protocol, the PBS feeds back the $\K$ smallest fading gains in the STPB channel to the SBS as well as the corresponding users indices. To avoid harmful interference at the PBS, the SBS schedules a SU  only if its STPB channel gain is made available at the SBS. 

This paper focuses on the effect of the $K$-SCG feedback protocol on the throughput scaling for three types of secondary networks: Total-Power-And-Interference-Limited (TPIL), Interference-Limited (IL) and Individual-Power-And-Interference-Limited (IPIL) networks when transmit powers of SUs are optimally allocated.  In the case of TPIL networks, transmit powers of SUs are limited by an average total power constraint as well as a constraint on the average total interference power that they cause to the PBS. On the other hand, transmit powers of SUs are limited by a constraint only on the average total interference power at the PBS for IL networks.  In the case of IPIL networks, transmit powers of SUs are limited by individual average power constraints as well as a constraint on the average total interference power at the PBS.  For each network type studied, we provide a sufficient condition on $\K$ such that the $K$-SCG feedback protocol is {\em asymptotically} optimal, {\em i.e.,} the sum-rate capacity scaling behavior under the $K$-SCG feedback protocol is the same with that under the full-feedback protocol.
 

Due to mathematical intractability of the cumulative distribution functions (CDF) of random variables emerging in secondary network capacity calculations, secondary network capacity scaling laws have been mainly investigated for specific fading distributions for STSB and STPB channel gains such as Rayleigh distribution in the literature, {\em e.g.,} see \cite{cogmud_twban09, cogmid_zhang10, cogmudmsd_10} and \cite{cogmud_alitajer10}.  Different from these works, this paper studies the sum-rate scaling behavior of secondary networks under both $K$-SCG and full-feedback protocols when distributions of STSB and STPB channel gains are arbitrarily chosen from a more general class of distribution functions called class $\cal C$-distributions ({\em i.e.,} see Definition \ref{Def1}).  The class $\cal C$-distributions contains distribution functions that decay double-exponentially and vary regularly around the origin.  It covers the most common fading distributions such as Rayleigh, Rician, Nakagami-$m$ and Weibull distributions.  In Appendix \ref{app: concentration}, we show that the concentration behavior of the extreme order statistic of an independent and identically distributed (i.i.d.) sequence of random variables is characterized by the asymptotic tail behavior of the CDF  common to all of them. This finding enables us to study the capacity scaling laws for secondary networks under class $\mathcal{C}$ distribution functions for STSB and STPB channel gains. 
 
Our results for the TPIL networks indicate that the secondary network throughput under the $K$-SCG and full-feedback protocols scales according to $\frac{1}{n_h}\log\log(\K )$ and $\frac{1}{n_h}\log\log(N )$, respectively. $n_h$ a is parameter determined from the asymptotic tail behavior of STSB channel power gains. For example, $n_h$ is equal to 1 for Rayleigh, Rician and Nakagami-$m$ distributions, whereas it is equal to $\frac{c}{2}$\footnotemark[1] \footnotetext[1]{$c$ is the Weibull fading parameter.} for the Weibull distribution. Consequently, for $\K =N^\delta$ where $0<\delta<1$, the secondary network throughput scales as $\frac{1}{n_h}\log\log(N)$.  Hence, the secondary network can achieve the same throughput scaling as with the full channel state information (CSI) case for any $\delta$ arbitrarily close to zero.  To put it in other words, the rate of growth of the feedback load can be made arbitrarily small when compared to the rate of growth of the number of SUs without any sacrifice from the optimal sum-rate scaling behavior.  

For $\K =\LO{N}$, we show that the interference power at the PBS converges to zero almost surely and also in mean as $N$ tends to infinity. From a practical point of view, this result implies that the interference  constraint cannot be satisfied with equality for $N$ large enough.  Hence, we can relax the interference constraint, {\em i.e.,} the average amount of total interference power at the PBS due to SU transmissions is not a performance limiting criterion any more. Once this happens, the SBS just requires the indices of the SUs for which $g_i\leq \g{\K }{N}$ rather than the actual realizations of the STPB channel gains, which further reduces the amount of feedback required between two networks. Furthermore, our results indicate that the sum-rate scaling behavior of TPIL networks is mainly affected by the distribution of STSB channel power gains rather than that of the STPB channel power gains.
 
In contrast to TPIL networks, the throughput scaling behavior of IL networks under $K$-SCG and full-feedback protocols is mainly affected by the distribution of STPB channel power gains rather than that of STSB channel power gains.  More specifically, our results for IL networks signify that the secondary network throughput under $K$-SCG and full-feedback protocols scales according to $\frac{1}{\gamma_g}\logp{N}$, where $\gamma_g$ is a parameter determined from the behavior of the CDF of the STPB channel power gains around the origin.  For example, $\gamma_g$ is equal to $1$, $m$ and $\frac{c}{2}$ for Rayleigh (as well as Rician), Nakagami-$m$ and Weibull distributions, respectively.  Hence, we conclude that the sum-rate scaling behavior in the IL networks is not affected by $\K$. That is, even for $\K=\BO{1}$, the sum-rate scaling behavior of an IL network under $K$-SCG and full-feedback protocols will be the same. This implies that the amount of required feedback between the PBS and SBS can be reduced substantially while keeping the same scaling behavior with that under the full-feedback protocol.   
 
Finally, our results for IPIL networks indicate that the secondary network throughput under both $K$-SCG and full-feedback protocols scales according to $\min\paren{1,\frac{1}{\gamma_g}}\logp{N}$. Similar to IL networks, $K$-SCG feedback protocol achieves the same scaling behavior as the full-feedback protocol does even for $\K=\BO{1}$, which again results in a substantial amount of feedback reduction at the SBS.  The throughput scaling behavior of IPIL networks under $K$-SCG and full-feedback protocols is mainly affected by the distribution of STPB channel power gains rather than that of STSB channel gains. Our results are summarized in Table \ref{Table: Main Results}.
 \begin{table}[!t]
\begin{minipage}{\textwidth}
\renewcommand{\arraystretch}{1.3}
\caption{Throughput Scaling behavior of $K$-SCG and Full-Feedback Protocols For Different Network Models }
\centering
\begin{tabular}{lll}
\toprule
\multicolumn{1}{c}{\multirow{2}{*}{ Network Model}} & \multicolumn{2}{c}{ Feedback Protocol}\\
 \cmidrule(r){2-3}

& \multicolumn{1}{c}{$K$-SCG} & \multicolumn{1}{c}{Full}  \\
\midrule
 Total-Power-And-Interference-Limited & $\lim\limits_{N\ra\infty}\frac{R_N\footnote{$R_N$ is the secondary network sum-rate.}}{\logp{\logp{\K}}}=\frac{1}{n_h}\footnote{ $n_h$ is parameter determined from the asymptotic tail behavior of the CDF of STSB channel power gains.}$ & $\lim\limits_{N\ra\infty}\frac{R_N}{\logp{\logp{N}}}=\frac{1}{n_h}$   \\
\midrule
 Interference-Limited & $\lim\limits_{N\ra\infty}\frac{R_N}{\logp{N}}=\frac{1}{\gamma_g}\footnote{$\gamma_g$ is a parameter determined from the behavior of the CDF of STPB channel power gains around the origin.}$ & $\lim\limits_{N\ra\infty}\frac{R_N}{\logp{N}}=\frac{1}{\gamma_g}$ \\
\midrule
Individual-Power-And-Interference-Limited & $\lim\limits_{N\ra\infty}\frac{R_N}{\logp{N}}=\min\paren{1,\frac{1}{\gamma_g}}$ & $\lim\limits_{N\ra\infty}\frac{R_N}{\logp{N}}=\min\paren{1,\frac{1}{\gamma_g}}$  \\
\bottomrule
\label{Table: Main Results}
\end{tabular}
\end{minipage}

\end{table}
  
\subsection{Related Work}
Jointly optimal power allocation and spectrum sharing in a cognitive radio setup with single SU has been extensively studied in the literature under different quality-of-service (QoS) criteria such as SU's outage probability and ergodic capacity, and under different constraints on transmit powers of SU and interference power at the primary receiver such as peak or average power and interference constraints \cite{Ghasemi07, Musavian09, Kang09}. These papers show that optimal resource allocation and interference management tasks in cognitive radio networks highly depends on the knowledge of secondary-transmitter-primary-receiver channel gain.  Optimal power allocation and spectrum sharing policy maximizing sum-rate in CMACs as well as cognitive broadcast channels (CBCs) under various transmit power and interference constraints has also been recently studied in \cite{RZhang09}.  It has been shown that the optimal power allocation for a CMAC under average transmit power and average interference constraints for continuous fading distributions is to schedule the SU with the best joint power and interference channel state. This result implies that the SBS requires interference channel gains of all SUs to perform the optimal scheduling and power allocation. Similar to these previous works, our performance measure in this paper is also the secondary network sum-rate capacity under the jointly optimal power control and spectrum sharing policy. Different from them, we focus on the feedback limited communication environments in which STPB channel states are available only for a subset of SUs at the SBS, and obtain tight sum-rate capacity scaling laws under such feedback limitations.  

Capacity scaling laws in CMACs under the complete knowledge of STPB channel gains has also been investigated in the literature, {\em e.g.,} see \cite{cogmud_twban09}, \cite{cogmid_zhang10} and \cite{NID12}, under various type of constraints on the transmit powers of SUs. The authors in \cite{cogmud_twban09} studied the capacity scaling laws for a multiple access secondary network for Rayleigh fading channels under joint peak transmit power and peak interference power constraints. They established logarithmic and double-logarithmic secondary network capacity scaling behavior under some approximations. Zhang \emph{et al.} \cite{cogmid_zhang10} extended these results to CMACs, CBCs, and cognitive parallel access channels. In \cite{NID12}, the authors studied throughput scaling behavior of IL and TPIL CMACs under full primary-secondary feedback assumption when transmit powers of SUs are optimally allocated. For specific communication environments, \emph{i.e.,} specific fading channel models for STSB and STPB channel gains, they showed that the secondary network sum-rate scales double logarithmically and logarithmically in TPIL and IL networks, respectively. These previous works did not consider feedback limited communication environments, and assumed very specific fading distributions to derive the stated sum-rate capacity scaling laws. 

Other related work includes secondary network capacity scaling in a multi-band setup such as \cite{cogmudmsd_10} and \cite{cogmud_alitajer10}. In \cite{cogmudmsd_10}, the authors studied the multiuser and multi-spectrum diversity gains for a cognitive broadcast network sharing multiple orthogonal frequency bands with a primary network.  Assuming Rayleigh fading channels, they analytically derived capacity expressions for the secondary network when the transmit power at each band is limited by a constraint on the peak interference power that the SBS can cause to the primary network. In \cite{cogmud_alitajer10}, the authors considered $N$ secondary transmitter-receiver pairs sharing $M$ frequency bands with a primary network. Under the optimum matching of $M$ SUs with $M$ primary network frequency bands, they obtained a double-logarithmic scaling law for the secondary network capacity for Rayleigh fading channels.  Although the problem formulation in the current paper is different than that in these previous works, similar techniques as in \cite{orderstat} are used to derive capacity scaling laws. We believe some parts of our analysis are expected to find greater applicability to extend sum-rate capacity scaling laws obtained for the dual broadcast channels with multiple transmission bands beyond Rayleigh fading communication environments. 

Finally, this work is also partially related to the cooperative multiple access channels (CO-MACs) in which ideal (error-free and infinite-capacity) backhaul links were originally considered to convey the received signal from each base-station to a remote central processor that performs joint data decoding.  Effect of finite-capacity backhaul links on the capacity of CO-MACs has been studied in the literature, and different multi-cell processing protocols has been proposed to cope with the backhaul link resource limitations, \emph{e.g.,} see \cite{Simeone09}, \cite{Sanderovich09} and references therein. From an engineering point of view, cognitive radio network planning task, \emph{e.g.,} design of primary-secondary backhaul links and efficient primary-secondary feedback protocols, highly depends on the knowledge of the secondary network capacity limitations under resource limited primary-secondary backhaul links.  However, it should be noted that the problem here is fundamentally different from CO-MACs in that SBS and PBS separately perform the signal decoding, and the backhaul link is only used to convey STPB channel gains to the SBS rather than the PBS received signal.

\subsection{ A Note on Notation and Paper Organization}
When we write $p(x)=\BO{q(x)}$ and $p(x)=\LO{q(x)}$ for two positive functions $p(x)$ and $q(x)$, we mean $\limsup_{x \ra \infty}\frac{p(x)}{q(x)}<\infty$ and $\lim_{x \ra \infty}\frac{p(x)}{q(x)} = 0$, respectively. By $p(x)=\BT{q(x)}$, we mean  $\limsup_{\ra\infty}\frac{p(x)}{q(x)}<\infty$ $\liminf_{x\ra\infty}\frac{p(x)}{q(x)}>0$.

As is standard in the literature \cite{Simon-Alouini05}, when we say a wireless channel is Rayleigh fading channel, we mean the channel magnitude gain is Rayleigh distributed, or equivalently the channel power gain is exponentially distributed.  By a Rician-$K_f$ fading channel, we mean the channel magnitude gain is Rician distributed with a Rician factor $K_f$.  For a Rician-$K_f$ fading channel, the channel power gain is non-central chi-square distributed with two degrees of freedom \cite{Stuber96}. When we say a wireless channel is Nakagami-$m$ distributed, we mean the channel magnitude gain is Nakagami distributed with a Nakagami factor $m \geq 0.5$. For a Nakagami-$m$ fading channel, the channel power gain is Gamma distributed. By a Weibull fading channel, we mean the channel magnitude gain is Weibull distributed with parameter $c>0$.  We refer the reader to \cite{Simon-Alouini05}, \cite{Stuber96} and \cite{Bertoni88} for more details about fading distributions.

The rest of the paper is organized as follows. Section \ref{Sec: System Model} describes the system model and network configuration along with our modeling assumptions. Section \ref{Sec: R&D} derives and compares the secondary network sum-rate scaling under $K$-SCG and full feedback protocols,  discusses the effect of fading channel parameters on the scaling laws, provides various insights into the derived throughput scaling laws, and illustrates the accuracy of our results by means of numerical study for cognitive radio networks with finitely many SUs.  Section \ref{sec: conclusion} concludes the paper. All proofs are relegated to the Appendix. 

 \section{System Model, Operating Constraints and the Network Types}\label{Sec: System Model}
 In this section, we will introduce the details of our system model, the operating constraints on the cognitive radio environment that go with this model and the classification of the network types studied throughout the paper based on these operating constraints. 

\subsection{System Model} 
\begin{figure}[!t]
\centering{\includegraphics[scale=0.6]{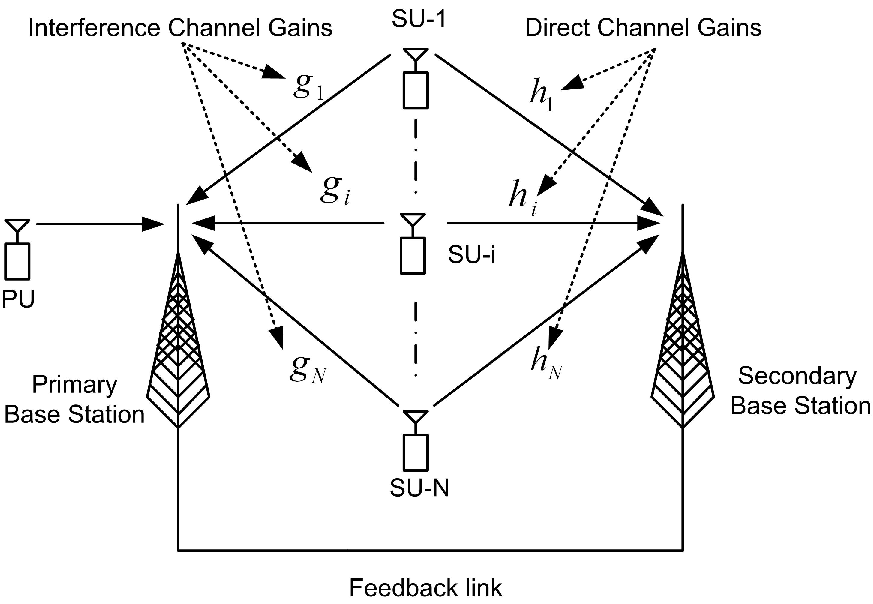}}
\caption{$N$ SUs forming a multiple access channel to the SBS and interfering with signal reception at the PBS. The backhaul feedback link can be implemented by using either a microwave link or a DSL link with limited capacity.} \label{F1}
\end{figure}

We consider an underlay cognitive uplink in which $N$ SUs transmit data to an SBS and interfere with the signal reception at a PBS.  Let $h_i$ and $g_i$ represent the fading power gains for the $i$th {\em direct} and {\em interference} links, respectively.  The classical ergodic block fading model \cite{Tse} is assumed to hold to model statistical variations in channel states for all direct and interference links. Further, we assume that $h_i$'s are independent and identically distributed (i.i.d.) random variables among themselves and $g_i$'s are i.i.d. random variables among themselves, but direct channel gains $h_i, i=1, \ldots, N$ may have a different joint distribution than that of interference channel gains $g_i$, $i=1, \ldots, N$. That is, the  random vectors ${\mathbi h}=\sqparen{h_1,h_2,\ldots,h_N}^\top$ and ${\mathbi g}= \sqparen{g_1,g_2,\ldots,g_N}^\top$ are also independent, but possibly with different distributions.  The explained communication set-up is represented in Fig. \ref{F1} pictorially.

To describe the direct and interference channel variations over time, we consider a general class of parametrized distributions, which is formally introduced in the following definition.   
\begin{definition}\label{Def1}
We say that the CDF of a random variable $X$, denoted by $F_X$, belongs to the class-$\cal C$ distributions if it satisfies the following properties:
\begin{itemize}
\item $F_X\paren{x}$ is continuous.
\item $F_X(x)$ has a positive support, {\em i.e.,} $F(x)=0$ for $x \leq 0$.
\item $F_X(x)$ is strictly increasing. 
\item The tail function $1-F(x)$ decays to zero \emph{exponentially}, {\em i.e.,} there exist constants $\alpha>0$, $\beta>0$, $n>0$, $l \in \R$ and a slowly varying function $H(x)$ satisfying $H(x) = \LO{x^{n}}$ such that
 $ \label{Eq: tail-condition-1}
 \lim_{x\ra\infty}\frac{1-F(x)}{\alpha x^{l}\e{\paren{-\beta x^{n}+H(x)}}}=1.\nonumber
$
\item $F(x)$ varies \emph{regularly} around the origin, {\em i.e.,} there exist constants $\eta>0$ and $\gamma>0$ such that
  $\label{Eq: tail-condition-2}
 \lim_{x \ra 0}\frac{F(x)}{\eta x^{\gamma}}=1.\nonumber
$
\end{itemize}
  \end{definition} 

Our results in Theorems \ref{Theo: IL}, \ref{Theo: IPIL} and \ref{Theo: TPIL} in Section \ref{Sec: R&D} indicate that the channel gain distribution parameters play important roles in identifying the pre-log factor in the fundamental capacity scaling laws for cognitive radio networks.  In particular, the decay rate of the CDF around zero and that of its associated tail function around infinity determine the nature of {\em full} MDG, which is otherwise hidden by only considering the Rayleigh fading scenario.  We will elaborate on these findings further as we discuss the above theorems in Section \ref{Sec: R&D}.  The parameters characterizing the behavior of the distribution of fading power gains around zero and infinity are illustrated in Table \ref{Table: Fading Parameters} for the commonly used fading models in the literature. To avoid any confusion, we represent these parameters with subscript $h$ for direct channel gains and with subscript $g$ for interference channel gains, \emph{e.g.,} $\eta_g$ or $\eta_h$, in the remainder of the paper.

\begin{table*}[!t]
\begin{minipage}{\textwidth}
\renewcommand{\arraystretch}{0.7}
\caption{Common fading channel models and their parameters}
\centering
\begin{tabular}{cccccccc}
\toprule
\multicolumn{1}{c}{\multirow{2}{*}{ Channel Model}}  & \multicolumn{7}{c}{ Parameters }\\
 \cmidrule(r){2-8}
&  $\alpha$ & $l$& $\beta$ & $n$ & $H(x)$ & $\eta$ & $\gamma$  \\
\midrule
Rayleigh & 1&  $0$ & 1 & 1 & 0 & 1&1 \\
\midrule
Rician-$k$ & $\frac{1}{2\sqrt{\pi}\e{k}\sqrt[4]{k\paren{k+1}}}$ & $-\frac{1}{4}$ & $k+1$ &1 & $2\sqrt{k\paren{k+1}x}$& $\frac{k+1}{\e{k}}$ & 1\\
\midrule
Nakagami-$m$ & $\frac{m^{m-1}}{\Gamma(m)}$ & $m-1$ & $m$ & 1 & 0& $\frac{m^{m-1}}{\Gamma(m)}$& $m$ \\
\midrule
Weibull-$c$ & 1& 0 & $\Gamma^\frac{c}{2}\paren{1+\frac{2}{c}}$ & $\frac{c}{2}$ & $0$ & $\Gamma^\frac{c}{2}\paren{1+\frac{2}{c}}$ & $\frac{c}{2}$ \\\bottomrule
\label{Table: Fading Parameters}
\end{tabular}
\end{minipage}
\end{table*}

\subsection{Operating Constraints}
We consider different operating constraints on the cognitive radio environment introduced above in order to identify the network types studied throughout the paper more systematically.  Specifically, different constraints on the transmission powers of SUs and the capacity of the feedback link are considered.  In each case, we analyze the throughput scaling behavior of the secondary network when the transmission powers of SUs are allocated according to an {\em optimum} power allocation policy subject to these constraints, where we define a power allocation policy $\PV=\left [\PS{1},\cdots,\PS{N}\right ]^\top$ as a mapping from $\R^{2N}$ to $\R^N$ in which $\PS{i}$ is the transmission power of the $i$th SU.

The equations  \eqref{TPC}, \eqref{IPC}, \eqref{IC},  \eqref{FC} and \eqref{PSC} below list the operating constraints on the studied cognitive radio environment formally, and the throughput scaling behavior of the secondary networks are derived systematically under different combinations of these constraints. 
\begin{subnumcases}{}
 \EW_{{\mathbi h},\mathbi {g}}\left[\mathbf 1^\top\PV\right]\leq P_{ \rm ave}, \label{TPC} \\
  \EW_{{\mathbi h},\mathbi {g}}\left[\PS{i}\right]\leq P_{ \rm ave}\quad 1\leq i\leq N, \label{IPC}\\
\EW_{{\mathbi h},\mathbi{g}}\left[\mathbi g^\top\PV\right]\leq Q_{ \rm ave}, \label{IC}\\
\PS{i}\I{g_i \geq \g{\K}{N}}= 0\quad  1\leq i\leq N, \label{FC}\\
\PS{i}\geq0 \label{PSC}.
\end{subnumcases}
Above, \eqref{TPC}, \eqref{IPC} and \eqref{IC} are average total power, average individual power and average total interference power constrains, respectively. \eqref{FC} is a constraint to guarantee that a SU is allowed to transmit {\em only if} its interference channel gain is available at the SBS, where $\g{\K }{N}$ represents the $\K$th smallest value in the set $\left\{{g_i}\right\}_{i=1}^N$.  \eqref{FC} will be called the {\em feedback constraint} in the remainder of the paper  as it describes the SUs that are allowed for transmission as a function of the feedback load $K_N$. Equation \eqref{PSC} is the usual positivity constraint on the transmission power, which is added for the sake of mathematical completeness.\footnote{In addition to feedback, another important issue in this setup is the estimation of interference channel gains by the PBS.  This can be efficiently done by using pilot signals transmitted intermittently by SUs. These pilot signals are heard by the PBS through interference channels, which can be further utilized to estimate interference channel gains.}

\subsection{TPIL Networks:}
In TPIL networks, we examine the secondary network throughput scaling behavior in two communication scenarios (CoSs) of interest: $\CoSTPILFull$ and $\CoSTPILRed$. $\CoSTPILFull$ refers to a TPIL network under full cooperation scenario whereas $\CoSTPILRed$ refers to a TPIL network under $K$-out-of-$N$ feedback protocol. In $\CoSTPILFull$, transmission powers of SUs are limited by an average total power constraint and an average total interference power constraint without any restriction on the amount of feedback information to be exchanged between the PBS and SBS.  Hence, transmission powers of SUs are allocated according to the solution of the following optimization problem: 
  \begin{eqnarray}\label{Prob: TPI-Full}
 \begin{array}{lcr}
\max\limits_{{\PV}}\EW_{{\mathbi h},\mathbi{g}}\left[\log\paren{1+\frac{\mathbi h^\top\PV}{W}}\right]\\
\text{subject to : \eqref{TPC}, \eqref{IC} and \eqref{PSC}}
  \end{array},
 \end{eqnarray}
where $W=N_0 + I$ represents the {\em average} background noise plus primary interference power at the SBS, whose realizations are modelled as a circularly symmetric complex Gaussian random variable with {\em slowly} varying variance.\footnote{If primary interference power changes at a time scale comparable to the fading process, our throughput scaling results in Section \ref{Sec: R&D} should be thought to hold point-wise for each realization of $W$.} It is important to note here that our control actions are only on the SUs since we consider the underlay communication paradigm for the co-existence of primary and secondary networks \cite{Goldsmith09}. Hence, the uncontrolled variable $W$ just functions as a scaling parameter for the secondary network direct channel gains.
\footnote{The joint control of primary and secondary networks is outside the scope of this paper. This will require a central authority that can oversee all control variables and channel gains, which implies excessive feedback load between primary and secondary networks and does not serve the purpose of throughput analysis under limited primary-secondary coordination. Further, we can also assume that the SBS can cancel primary interference to remove the coupling between primary and secondary networks as in [7].}     

The solution to \eqref{Prob: TPI-Full} was given in \cite{NID12} as in \eqref{Eq: powalloc}.
\begin{figure*}
 \begin{eqnarray} 
\OPC{i}{N} =
\left\{
\begin{array}{cc}
 \paren{\frac{1}{\lambda_N + \mu_N g_i} -  \frac{W}{h_i}}^{+} & \mbox{if }  i=\arg\max\limits_{1\leq j\leq N }\frac{h_j}{\lambda_N+\mu_Ng_j} \\
 0 & \mbox{otherwise}
\end{array}.
\right. \label{Eq: powalloc}
\end{eqnarray}
\hrule
\end{figure*}
This result intuitively indicates that the jointly optimum spectrum sharing and power control policy maximizing information theoretic throughput capacity of a cognitive uplink with full primary cooperation under average total transmission and interference power constraints is to schedule the SU with the best joint direct and interference channel state summarized by the random variable $\X{\lambda_N}{\mu_N}=\max_{1\leq i\leq N}\frac{h_i/W}{\lambda_N+\mu_N g_i}$ according to a power allocation policy in the form of a water-filling algorithm with changing water levels.  Here, $\lambda_N$ and $\mu_N$ are Lagrange multipliers associated with the average total transmission and interference power constraints, respectively. We note that there is no ambiguity with the solution described in \eqref{Eq: powalloc} since direct and interference channel gains are continuous random variables, and there is only one SU achieving the maximum joint channel state with probability one. Let $\RateTPILFull$ be the throughput of the secondary network for the {\em all} feedback scenario.  Then, it follows directly that
\begin{eqnarray}
\RateTPILFull =\ES{\logp{\X{\lambda_N}{\mu_N}}\I{\X{\lambda_N}{\mu_N}\geq1}}. \nonumber
\end{eqnarray}

Different from the full primary cooperation scenario, transmission powers of SUs in $\CoSTPILRed$ are also limited by an extra feedback constraint given by \eqref{FC}, besides the average total transmission and interference power constraints above.  Hence, transmission powers of SUs in this case are allocated according to the solution of the following optimization problem:
  \begin{eqnarray}\label{Prob: TPI-Red}
 \begin{array}{lcr}
\max\limits_{{\PV}}\EW_{{\mathbi h},\mathbi{g}}\left[\log\paren{1+ \frac{\mathbi h^\top\PV}{W}}\right]\\
\text{subject to : \eqref{TPC}, \eqref{IC}, \eqref{FC}, and \eqref{PSC}}
  \end{array}.
 \end{eqnarray}
 \begin{lemma}
Let $\pi(j)$ be a mapping from $\left\{1,\cdots,\K\right\}$ to $\left\{1,\cdots,N\right\}$ such that
\begin{eqnarray}
\pi(j)=i \quad \text{if}\quad g_i=\g{j}{N}.\nonumber
\end{eqnarray}
Then, the solution for \eqref{Prob: TPI-Red} is given by \eqref{Eq: powallocP1}.
\begin{figure*}
 \begin{eqnarray}\label{Eq: powallocP1}
\OPC{i}{\K} =
\left\{
\begin{array}{cc}
 \paren{\frac{1}{\lambda_N + \mu_N g_i} -  \frac{W}{h_i}}^{+} & \mbox{if }  i=\pi\paren{\arg\max\limits_{1\leq j\leq \K }{\frac{h_{\pi(j)}}{\lambda_N+\mu_Ng_{\pi(j)}}}}  \\
 0 & \mbox{otherwise}
\end{array}.
\right.
\end{eqnarray}
\hrule
\end{figure*}
 \end{lemma}
\begin{IEEEproof}
Follows directly by inspecting the structure of the solution given for \eqref{Prob: TPI-Full} in \eqref{Eq: powalloc}.
\end{IEEEproof}

As an analogy with the solution described in \eqref{Eq: powalloc}, the jointly optimum spectrum sharing and power allocation policy described in \eqref{Eq: powallocP1} under the limited primary cooperation is to schedule the SU with the best joint channel state among the ones that are fed back to the SBS. Specifically, the throughput in the $\CoSTPILRed$ scenario can be written as
\begin{eqnarray}\label{Eq: RateTPIL}
\RateTPILRed=\ES{\log(\Xr{\lambda_N}{\mu_N})\I{\Xr{\lambda_N}{\mu_N}\geq1}},\nonumber
\end{eqnarray}
where $\Xr{\lambda_N}{\mu_N}=\max_{1\leq j\leq \K }\frac{h_{\pi(j)}/W}{\lambda_N+\mu_Ng_{\pi(j)}}$.  This expression makes it further clear that the jointly optimum spectrum sharing and power control policy maximizing information theoretic throughput capacity of a cognitive uplink under the $K$-out-of-$N$ feedback protocol with average total transmission and interference power constraints is to schedule the SU with the best joint direct and interference channel state among the ones whose interference channel states are fed back to the SBS.

\subsection{IL Networks:}
We study the throughput scaling behavior of IL networks under two CoSs of interest: $\CoSILFull$ and $\CoSILRed$. In $\CoSILFull$, transmission powers of SUs are limited only by an average total interference power constraint.  In this case, the secondary network throughput is given by:
\begin{eqnarray}
\RateILFull =\ES{\logp{\X{0}{\mu_N}}\I{\X{0}{\mu_N}\geq1}}. \nonumber
\end{eqnarray}
In addition to the average total interference power constraint, transmission powers of SUs are also limited by the feedback constraint \eqref{FC} in $\CoSILRed$.  The secondary network throughput in $\CoSILRed$ is given by:
\begin{eqnarray}
\RateILRed =\ES{\logp{\Xr{0}{\mu_N}}\I{\Xr{0}{\mu_N}\geq1}}, \nonumber
\end{eqnarray}
where the random variable $X_{K_N}^\star\paren{\lambda_N, \mu_N}$ summarizing the best joint channel state under {\em limited} primary cooperation defined as above.
\subsection{IPIL Networks:}
Similar to the above cases, we investigate the secondary network throughput scaling behavior of IPIL networks under two CoSs of interest: $\CoSIPILFull$ and $\CoSIPILRed$.  In $\CoSIPILFull$, transmission powers of SUs are limited by individual average transmission power constraints and an average total interference power constraint.  Hence, transmission powers of SUs are allocated according to the solution of the following optimization problem:
\begin{eqnarray}\label{Prob: IPI-Full}
\begin{array}{lcr}
\max\limits_{{\PV}}\EW_{{\mathbi h},\mathbi{g}}\left[\log\paren{1+\frac{\mathbi h^\top\PV}{W}}\right]\\
\text{subject to : \eqref{IPC}, \eqref{IC} and \eqref{PSC}}
\end{array}.\nonumber
\end{eqnarray}
The throughput in $\CoSIPILFull$ is given by:
\begin{eqnarray}
\RateIPILFull =\ES{\logp{\X{\lambda_N}{\mu_N}}\I{\X{\lambda_N}{\mu_N}\geq1}}, \nonumber
\end{eqnarray}
where $\lambda_N$ now represents the Lagrange multiplier associated with individual transmission power constraints. Again, the definition of $X_N^\star\paren{\lambda_N, \mu_N}$ is the same with the one above, except with a change of interpretation of the Lagrange multiplier $\lambda_N$ in this network.  Hence, although the functional structure of the power control policy is the same for both cases of TPIL and IPIL networks, the resulting transmission powers can be much different.  In the first case, $\lambda_N$ is chosen to keep the aggregate transmission power around $P_{ \rm ave}$ whenever there is a transmission from the totality of all SUs. Hence, each transmission is expected to occur with power around $P_{\rm ave}$ in TPIL networks. On the other hand,  $\lambda_N$ is chosen to keep individual transmission powers around $P_{ \rm ave}$ in IPIL networks. Therefore, considering the spectrum access probability, each transmission is expected to occur with power around $P_{\rm ave}$ times the probability of being scheduled for transmission in the second case.  This difference in turn results in different throughput scaling behavior for both networks as explained in detail in Section \ref{Sec: R&D}.      

In $\CoSIPILRed$, in addition to the individual average transmission power and average total interference power constraints, transmission powers of SUs are also limited by the feedback constraint in \eqref{FC}. Hence, the transmission powers of SUs are allocated according to the solution of the following optimization problem:
 \begin{eqnarray}\label{Prob: IPI-Red}
\begin{array}{lcr}
\max\limits_{{\PV}}\EW_{{\mathbi h},\mathbi{g}}\left[\log\paren{1+ \frac{\mathbi h^\top\PV}{W}}\right]\\
\text{subject to : \eqref{IPC}, \eqref{IC}, \eqref{FC} and \eqref{PSC}}
\end{array}.\nonumber
\end{eqnarray}

The throughput in $\CoSIPILRed$ is given by:
\begin{eqnarray}
\RateIPILRed =\ES{\logp{\Xr{\lambda_N}{\mu_N}}\I{\Xr{\lambda_N}{\mu_N}\geq1}}.\nonumber
\end{eqnarray}
Again, the definition of $X_{K_N}^\star\paren{\lambda_N, \mu_N}$ is the same with the one given above, except with a slight change of interpretation of the Lagrange multiplier $\lambda_N$ in this network type.

\begin{remark}
Although, the same notations $\lambda_N$ and $\mu_N$ are used to represent the Lagrange multipliers for different network types, their association to the constraints will be clear from the context. In particular, $\lambda_N$ will represent the Lagrange multiplier associated with the total average transmission power constraint in $\CoSTPILFull$ and $\CoSTPILRed$, whereas it will represent the identical Lagrange multipliers associated with individual average transmission power constraints in $\CoSIPILFull$ and $\CoSIPILRed$ in the remainder of the paper. Also, $\mu_N$ represents the Lagrange multiplier associated with the average interference power constraint in all CoSs.
\end{remark}

\section{Results and Discussions}\label{Sec: R&D}
In this section, we state the main asymptotic sum-rate scaling results of the paper along with numerical analysis illustrating them for finite networks.  We also discuss various insights about the derived sum-rate scaling results.  The proofs are relegated to the appendices for the sake of fluency of the paper. Our first result establishes the scaling behavior for $\RateTPILFull$ and $\RateTPILRed$. 
\begin{theorem}\label{Theo: TPIL}
Let $\K$ grow to infinity at a rate $\K=\LO{N}$.  Then, the sum-rates $\RateTPILFull$ and $\RateTPILRed$ under $\CoSTPILFull$ and $\CoSTPILRed$, respectively, scale according to  
\begin{eqnarray}
\lim_{N\ra\infty}\frac{\RateTPILRed}{\logp{\logp{\K}}}=\lim_{N\ra\infty}\frac{\RateTPILFull}{\logp{\logp{N}}}=\frac{1}{n_{h}}.\nonumber
\end{eqnarray}
\end{theorem}
\begin{IEEEproof}
Please see Appendix \ref{App: TPIL}.
\end{IEEEproof}

 In Appendix \ref{App: TPIL}, we give a detailed proof for Theorem \ref{Theo: TPIL} for $\CoSTPILRed$, and only the key proof ideas for $\CoSTPILFull$ are illustrated to avoid repetition. 
 Theorem \ref{Theo: TPIL} indicates that the secondary network throughput scales double-logarithmically under $\CoSTPILFull$ and $\CoSTPILRed$ with $N$ and $\K$, respectively, when distributions of STPB and STSB channel power gains belong to class $\mathcal{C}$-distributions. Hence, for $\K=N^\delta$ and $\delta\in\left(0,1\right)$, the secondary network throughput scaling behavior under $\CoSTPILFull$ and $\CoSTPILRed$ are the same.  Since $\delta$ can be chosen arbitrarily close to zero, this result implies that under $K$-SCG feedback protocol, the amount of feedback in the PSFL can be dramatically reduced while the secondary network still achieves the same scaling behavior as the one achieved by the full-feedback protocol. 
 
The dependence of multiuser diversity gains (MDGs) in $\CoSTPILRed$ on $\K$ indicates that the STSB channel gains are the major source of MDGs in TPIL networks. This is mainly because the Lagrange multipliers $\lambda_N$ cannot be made arbitrarily close to zero (see Lemma \ref{Lem: Lambda-Convergence} in appendix \ref{App: TPIL}) in this case, and as a result, the asymptotic behavior of $\max_{1\leq i\leq \K}\frac{h_{\pi(i)}}{\lambda_N+\mu_N g_{\pi(i)}}$ is primarily governed by the distribution of STSB channel gains.  Larger $\K$ implies that more STPB channel gains are available at the SBS, and we observe a corresponding increase in the MDG. Moreover, the Theorem \ref{Theo: TPIL} reveals that the secondary network sum-rate scaling behavior under $\CoSTPILFull$ and $\CoSTPILRed$ is controlled by a pre-log factor of $\frac{1}{n_h}$. To put it another way, the available degrees of freedom for the cognitive multiple access channel in question reversely depends on the tail decay rate of the CDF of the STSB channel power gains\footnote{Note that for distribution functions belonging to the class $\mathcal{C}$-distributions, $n$ mainly controls their tail decay rates (see Definition \ref{Def1}).}. The pre-log factor is equal to $\frac{2}{c}$ for the Weibull distributed STSB channel gains, and equal to $1$ for Rayleigh, Rician and Nakagami-$m$ distributed STSB channel gains.

In Appendix \ref{App: TPIL}, we show that the Lagrange multipliers $\lambda_N$ converge to $\frac{1}{P_{\rm ave}}$ as $N$ becomes large both in $\CoSTPILFull$ and $\CoSTPILRed$. This finding is helpful to study the second order effects of the average total power constraint $P_{\rm ave}$ on the secondary network throughput under $\CoSTPILFull$ and $\CoSTPILRed$.  Based on our analysis in Appendix \ref{App: TPIL}, we characterize the second order effects of $P_{\rm ave}$ and other fading parameters on the secondary network throughput under $\CoSTPILFull$ for finite number of SUs by bounding $\RateTPILFull$ from above and below as 
\begin{eqnarray}\label{Eq: TPILFull-up-low}
\lefteqn{\paren{1-\epsilon}\frac{1}{n_h}\log\logp{N}+\logp{P_{\rm ave}}+\frac{1}{n_h}\logp{\frac{1}{\beta_h}}+\BO{1}\leq \RateTPILFull\leq}\hspace{ 16cm}\nonumber\\
\lefteqn{  \paren{1+\epsilon}\frac{1}{n_h}\log\logp{N}+\logp{P_{\rm ave}}+\frac{1}{n_h}\logp{\frac{1}{\beta_h}}+\BO{1},}\hspace{ 10cm}
 \end{eqnarray} 
 for all $\epsilon>0$ and $N$ large enough ({\em i.e.,} see \eqref{Eq: FinalTPILFull}). Therefore, an increase in $P_{\rm ave}$ results in a corresponding logarithmic increase in $\RateTPILFull$, implying that $P_{\rm ave}$ has a logarithmic effect on $\RateTPILFull$. For a given a fading model for STSB channel gains, the constant term $\frac{1}{n_h}\logp{\frac{1}{\beta_h}}$ in \eqref{Eq: TPILFull-up-low} can be thought of being the second order effect of the fading model on $\RateTPILFull$ for finitely many SUs.  $\frac{1}{n_h}\logp{\frac{1}{\beta_h}}$ is equal to $\logp{\frac{1}{K_f+1}}$ for the Rician distributed STSB channel gains, and equal to $\logp{\frac{1}{m}}$ for the Nakagami-$m$ distributed STSB channel gains. This implies that for a fixed number of SUs, as the Rician factor $K_f$ or the Nakagami-$m$ parameter $m$ increases, we observe a logarithmic reduction in the secondary network throughput.  The reason for this behavior is that STSB channel gains become more deterministic as $K_f$ or $m$ increases, and as a result, the MDG drops since it depends on the dynamic range of the CDF of the STSB channel gains. 
 
 For the Weibull distributed STSB channel gains, $\frac{1}{n_h}\logp{\frac{1}{\beta_h}}$ is equal to $\logp{\frac{1}{\Gamma\paren{1+\frac{2}{c}}}}$, which first increases and then decreases as the Weibull fading parameter $c$ grows large. This behavior can be explained as follows.  For small values of $c$, the Weibull distribution is concentrated around zero, \emph{i.e.,} it is almost deterministic, whereas its dynamic range expands as $c$ increases. Thus, the second order term $\logp{\frac{1}{\Gamma\paren{1+\frac{2}{c}}}}$ in sum-rate expression increases as $c$ increases from zero.  On the other hand, as $c$ becomes large, after a certain point, the Weibull distribution starts to concentrate around one, \emph{i.e.,} it becomes deterministic again, and as a result the second order term $\logp{\frac{1}{\Gamma\paren{1+\frac{2}{c}}}}$ drops again.  Finally, we note that $P_{\rm ave}$ and STSB fading distribution parameters have the similar logarithmic second order effects on the secondary network throughput under $\CoSTPILRed$ ({\em i.e.,} see \eqref{Eq: FinalTPILRed}).
 
\begin{figure}[t]
\centering
\subfigure[]
{
\begin{tikzpicture}
\begin{axis}[title={Throughput in $\CoSTPILFull$ and $\CoSTPILRed$},legend style={yshift=-3.6cm, xshift=0.05cm},\axissetting]
                                      
\pfgsetting
\textbox{4.6cm}{1.7cm}{3.2cm}{Weibull, $c=4$}{Nakagami, $m=0.5$}
\addplot+[color=blue,mark=none] table [x=N,y=0.5log]{WNTPIL1.dat};\addlegendentry{$0.5\log\logp{N}+\logp{P_{\rm ave}}+\frac{1}{n_h}\logp{\frac{1}{\beta_h}}$};
\addplot+[color=red,mark=square] table [x=N,y=f]{WNTPIL1.dat};\addlegendentry{$\K=N$};
\addplot+[mark=o, mark size=2.5pt] table [x=N,y=K0.8]{WNTPIL1.dat};\addlegendentry{$\K=N^{0.8}$};
\end{axis}
\end{tikzpicture}
\label{WNTPIL1}}
\subfigure[]
{
\begin{tikzpicture}
\begin{axis}[title={Throughput in $\CoSTPILFull$ and $\CoSTPILRed$},legend style={yshift=-3.6cm, xshift=0.05cm},\axissetting]
\pfgsetting
\textbox{4.6cm}{1.7cm}{3.2cm}{Weibull, $c=1$}{Nakagami, $m=0.5$}
\addplot+[color=blue,mark=none] table [x=N,y=2log]{WNTPIL2.dat};\addlegendentry{$2\log\logp{N}+\logp{P_{\rm ave}}+\frac{1}{n_h}\logp{\frac{1}{\beta_h}}$};
\addplot+[color=red,mark=square] table [x=N,y=f]{WNTPIL2.dat};\addlegendentry{$\K=N$};
\addplot+[mark=o, mark size=2.5pt] table [x=N,y=K0.8]{WNTPIL2.dat};\addlegendentry{$\K=N^{0.8}$};
\end{axis}
\end{tikzpicture}
\label{WNTPIL2}}
\subfigure[]
{
\begin{tikzpicture}
\begin{axis}[title={Throughput in $\CoSTPILFull$ and $\CoSTPILRed$},legend style={yshift=-3.6cm, xshift=0.05cm},\axissetting]
\pfgsetting
\textbox{4.9cm}{1.7cm}{2.6cm}{Rayleigh}{Weibull, $c=1$}
\addplot+[color=blue,mark=none] table [x=N,y=log]{RWTPIL.dat};\addlegendentry{$\log\logp{N}+\logp{P_{\rm ave}}+\frac{1}{n_h}\logp{\frac{1}{\beta_h}}$};
\addplot+[color=red,mark=square] table [x=N,y=f]{RWTPIL.dat};\addlegendentry{$\K=N$};
\addplot+[mark=o, mark size=2.5pt] table [x=N,y=K0.8]{RWTPIL.dat};\addlegendentry{$\K=N^{0.8}$};
\end{axis}
\end{tikzpicture}
\label{RWTPIL}}
\subfigure[]
{
\begin{tikzpicture}
\begin{axis}[title={Throughput in $\CoSTPILFull$, $N=50$ },legend style={yshift=-3.7cm, xshift=0.05cm},xmin=0.2,xmax=2,xtick={0.2,0.4,0.6,0.8,1,1.2,1.4,1.6,1.8,2},grid=major, tick style={color=black, major tick length={0.10 cm}}, grid style={line width= 0.75pt, densely dotted, color= black}]
\pgfplotsset{
width= 8cm, 
every axis/.append style={line width=1.2pt},
label style={font=\bf\scriptsize}, 
ylabel style={yshift=-0.8em},
xlabel={STSB-FM Parameter $(c,m,K_f)$},
ylabel={Throughput (nats per channel use)},
title style={font=\bf\scriptsize}, 
tick label style={font=\scriptsize,/pgf/number format/1000 sep={} },
tick style={ line width=1.5pt},
legend style={font=\bf\tiny,cells={anchor=west}}
}
\addplot+[color=blue,mark=asterisk] table [x=N1,y=W]{TPILvsParameter1.dat};\addlegendentry{\bf STPB-FM: Rayleigh, STSB-FM: Weibull $(c)$};
\addplot+[color=red,mark=none] table [x=N1,y=N]{TPILvsParameter2.dat};\addlegendentry{\bf STPB-FM: Rayleigh, STSB-FM: Nakagami $(m)$};
\addplot+[mark=none] table [x=N3,y=R]{TPILvsParameter1.dat};\addlegendentry{\bf STPB-FM: Rayleigh, STSB-FM: Rician $(K_f)$};

\end{axis}
\end{tikzpicture}
\label{TPILvsParameter}}
\caption{ Secondary network throughput scaling under $\CoSTPILFull$ ($\K=N$) and $\CoSTPILRed$ ($\K=N^{0.8}$) for different communication environments (a)-(c). Secondary network throughput under $\CoSTPILFull$ as a function of the STSB-FM parameter for $N=50$ (d). $P_{\rm ave}$ and $Q_{\rm ave}$ are set to 15dB and 0dB, respectively.}
\label{FTPIL}
\end{figure}
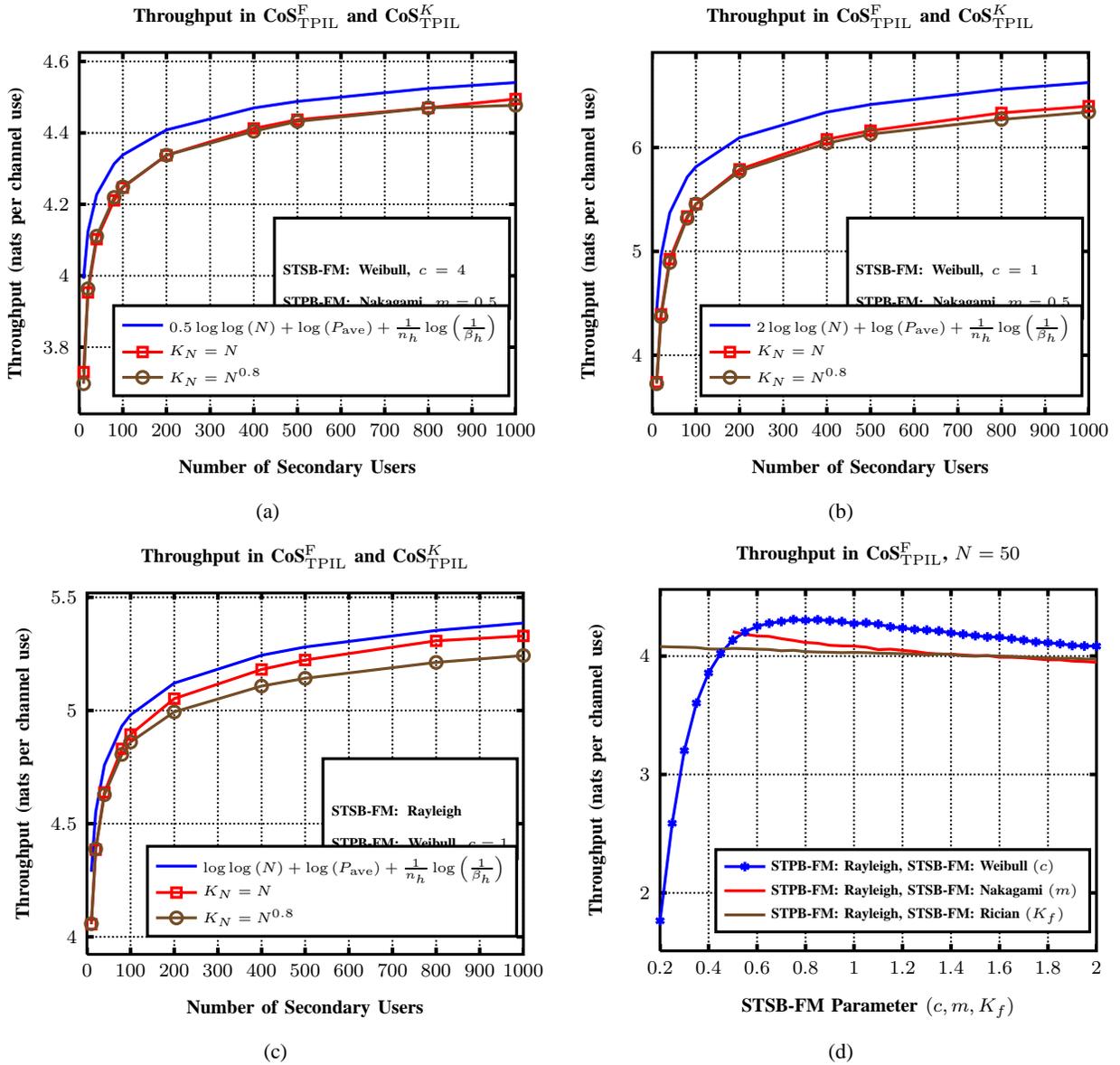

Fig. 2(a)-(c) demonstrate the sum-rate scaling behavior of the secondary network under $\CoSTPILFull$ and $\CoSTPILRed$ as a function of the number of SUs for different STSB-fading models (STSB-FMs) and STPB-fading models (STPB-FMs). $P_{\rm ave}$ and $Q_{\rm ave}$ are set to 15dB and 0dB, respectively. Similar qualitative behavior continues to hold for other values of $P_{\rm ave}$ and $Q_{\rm ave}$. In Fig. 2(a)-(c), the curves with $\K=N$ represent the secondary network sum-rate under $\CoSTPILFull$, and the curves with $\K=N^{0.8}$ represent the secondary network sum-rate under $\CoSTPILRed$. 

In Fig. \ref{WNTPIL1} and Fig. \ref{WNTPIL2}, STSB and STPB channel gains are distributed according to Weibull and Nakagami-$m$ fading models, respectively. Weibull fading parameter, $c$, is set to 4 in Fig. \ref{WNTPIL1} and to 1 in Fig. \ref{WNTPIL2}.  In both figures, Nakagami fading parameter $m$ is set to 0.5.  As Fig. \ref{WNTPIL1} and Fig. \ref{WNTPIL2} demonstrate, the sum-rate of the secondary network under $\CoSTPILFull$ and $\CoSTPILRed$ scales according to $\frac{2}{c}\log\logp{N}$ when STSB channel gains are Weibull distributed.  That is, the scaling behavior is $0.5\log\logp{N}$ for $c=4$ and $2\log\logp{N}$ for $c=1$, which is in accordance with the MDGs predicted by Theorem \ref{Theo: TPIL}.  Similar qualitative behavior continues to hold for other values of $m$ and $c$. 

In Fig. \ref{RWTPIL}, the Rayleigh fading model is used to model STSB channel variations, and the Weibull fading model is used to model STPB channel variations. The Weibull fading parameter $c$ is set to $1$.  As Fig. \ref{RWTPIL} shows, secondary network throughput scales according to $\log\logp{N}$ when STSB channel gains are Rayleigh distributed as predicted by Theorem \ref{Theo: TPIL}. Similar qualitative behavior continues to hold for other values of $c$. In particular, closeness of simulated data rates and $\frac{1}{n_h}\log\logp{N}+\logp{P_{\rm ave}}+\frac{1}{n_h}\logp{\frac{1}{\beta_h}}$ curves in Fig. \ref{FTPIL} further indicates the logarithmic effect of $P_{\rm ave}$ as well as other second order effects of the STPB fading parameters on the secondary network throughput under $\CoSTPILFull$ and $\CoSTPILRed$. Furthermore, as Fig. 2(a)-(c) show, throughput loss due to implementing the $K$-SCG feedback protocol is negligible, which indicates that the $K$-SCG feedback protocol is an effective primary-secondary feedback reduction policy even for finitely many SUs.

Fig. \ref{TPILvsParameter} shows the second order effects of the fading parameters on the secondary network sum-rate under $\CoSTPILFull$ for $N=50$.  In Fig. \ref{TPILvsParameter}, STPB channel gains are Rayleigh distributed, whereas Weibull, Nakagami and Rician fading models are considered for STSB channel gains.  As the Rician fading parameter $K_f$ or the Nakagami-$m$ fading parameter $m$ becomes large, the STSB channel gains become more deterministic, and as a result the secondary network throughput drops as predicted by our discussion above.  As the Weibull fading parameter $c$ becomes large, the secondary network throughput first increases and then decreases, which is also in accordance with our discussion above. 
\begin{remark}
In Appendix \ref{app: concentration}, we show that the concentration behavior of the extreme order statistic of an i.i.d. sequence of random variables with the common CDF $F(x)$, which does not have to have a closed form expression, is characterized by the functional inverse of the function $G(x)$ characterizing the tail behavior of $F(x)$, \emph{i.e.,} $\lim_{x\ra\infty}G(x)\paren{1-F(x)}=1$.  This is the key result used to establish the secondary network sum-rate scaling under different CoSs. 
\end{remark}
\begin{remark}
In Appendix \ref{App: TPIL}, we show that the sum-rate of a primary multiple access network with a total power constraint $\RateTPM$ scales according to $\lim_{N\ra\infty}\frac{\RateTPM}{\log\logp{N}}=\frac{1}{n_h}$ when the CDF of the channel gains belong to the class $\mathcal{C}$-distributions.  This result is used to establish the upper bound on the secondary network sum-rate under $\CoSTPILFull$.  
\end{remark}

Our next theorem establishes an important convergence behavior for the total interference power at the PBS under $\CoSTPILRed$ as the number of SUs grows large.
\begin{theorem}\label{Theo: Interference Convergence}
Let $\mathcal{I}_{\K}$ be the secondary network interference power at PBS under $\CoSTPILRed$. For $\K = \LO{N}$, $\lim_{N \ra \infty} \mathcal{I}_{\K} = 0$ almost surely and $\lim_{N \ra \infty} \ES{\mathcal{I}_{\K}} = 0$. 
\end{theorem}
\begin{IEEEproof}
Please see Appendix \ref{App: Interference-Conv}.
\end{IEEEproof}

In Appendix \ref{App: Interference-Conv}, we give a detailed proof for the convergence of $\mathcal{I}_{\K}$ to zero in mean, and then we use this result to conclude the almost sure convergence of $\mathcal{I}_{\K}$ to zero. These convergence results can be justified by the fact that $\g{\K }{N}$, \emph{i.e.,} the largest STPB channel gain available at the SBS under the $K$-SCG feedback protocol, converges to zero as $N$ becomes large for $\K=\LO{N}$. An important practical consequence of Theorems \ref{Theo: TPIL} and \ref{Theo: Interference Convergence} is that for $\K=N^\delta$ and $\delta\in\left(0,1\right)$, the secondary network under $\CoSTPILRed$ achieves the optimal throughput scaling behavior while the interference at the PBS becomes negligible as $N$ grows large. To put it in other words, the secondary network can co-exist with the primary network by {\em virtually} causing no interference, and yet still achieving the optimal data rates.  

It is also important to note that Theorem 2 implies the existence of a constant $N_0$ such that for all $N \geq N_0$, the average interference power constraint at the PBS cannot be satisfied with equality.  Therefore, the Lagrange multipliers associated with the average interference power constraint become zero for all $N$ large enough, {\em i.e.,} $\mu_N=0$ for all $N \geq N_0$.  As a result, the SBS just requires the index set $I_{\K} =\brparen{i: i = \pi(j), 1 \leq j \leq \K}$ to choose the SU with the best STSB channel gain for the optimum power allocation. From a practical point of view, this phenomenon provides an extra reduction in the total feedback load required to achieve the optimum throughput scaling for cognitive radio networks.  

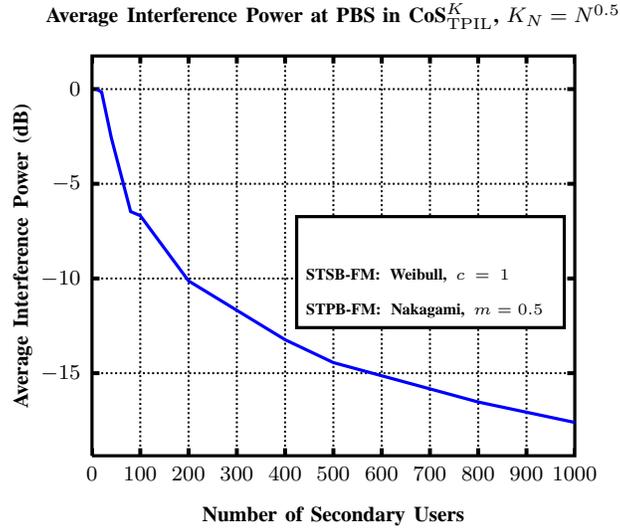
\begin{figure}[t]
\centering
\begin{tikzpicture}
\begin{axis}[title={Average Interference Power at PBS in $\CoSTPILRed$, $\K=N^{0.5}$},legend style={yshift=-3.5cm, xshift=0.12cm},\axissetting]
\pgfplotsset{
width= 8cm, 
every axis/.append style={line width=1.2pt},
label style={font=\bf\scriptsize}, 
ylabel style={yshift=-0.8em},
xlabel={Number of Secondary Users},
ylabel={Average Interference Power (dB)},
title style={font=\bf\scriptsize}, 
tick label style={font=\scriptsize,/pgf/number format/1000 sep={} },
tick style={ line width=1.5pt},
legend style={font=\bf\tiny,cells={anchor=west}}
}
\textbox{4.5cm}{2cm}{3.3cm}{Weibull, $c=1$}{Nakagami, $m=0.5$}
\addplot+[color=blue,mark=none] table [x=N,y=AveInt]{WNTPIL-Int.dat};
\end{axis}
\end{tikzpicture}
\caption{The change of average interference power at the PBS with $N$ under $\CoSTPILRed$. $\K=N^{0.5}$. $P_{\rm ave}$ and $Q_{\rm ave}$ are set to 15dB and 0dB, respectively.}
\label{AveInt}
\end{figure}

In Fig. \ref{AveInt}, we depict the average interference power at the PBS under $\CoSTPILRed$ for $\K=N^{0.5}$. $P_{\rm ave}$ and $Q_{\rm ave}$ are set to 15dB and 0dB, respectively.  The STSB channel gains are distributed according to the Weibull fading model with $c=1$, and the STPB channel gains are Nakagami-$m$ distributed with $m=0.5$.  As Fig. \ref{AveInt} shows, the average interference power at the PBS under $\CoSTPILRed$ converges to zero when the number of SUs becomes large, a behavior which was predicted by Theorem \ref{Theo: Interference Convergence}.
 
Our next theorem establishes the secondary network scaling behavior under $\CoSILFull$ and $\CoSILRed$. 
  \begin{theorem}\label{Theo: IL}
Let $\RateILFull$ and $\RateILRed$ be the secondary network throughput under $\CoSILFull$ and $\CoSILRed$ for $0<\K\leq N$, respectively. Then, 
\begin{eqnarray}
\lim_{N\ra\infty}\frac{\RateILFull}{\logp{N}}=\lim_{N\ra\infty}\frac{\RateILRed}{\logp{N}}=\frac{1}{\gamma_g}.\nonumber
\end{eqnarray}
\end{theorem}
\begin{IEEEproof}
Please see Appendix \ref{App: IL}.
\end{IEEEproof}

Theorem \ref{Theo: IL} establishes the logarithmic scaling behavior for the secondary network sum-rate with $N$ under $\CoSILFull$ and $\CoSILRed$ when the CDFs of STPB and STSB channel gains belong to the class $\mathcal{C}$-distributions.  Theorem \ref{Theo: IL} also indicates that the secondary network sum-rate scaling behavior under $\CoSILRed$ is independent of the scaling behavior of $\K$ with $N$.  Hence, the optimal secondary network throughput scaling behavior in $\CoSILRed$ can be attained even with $\K=\BO{1}$.  This is primarily because the STPB channel gains turn out to be the main source of MDGs in IL networks since $\frac{h_i}{g_i}$ and $\frac{1}{g_i}$ have the similar tail behavior {\em i.e.,} see Appendix \ref{App: IL}. Thus, the common CDF of the STPB channel gains characterizes the asymptotic behavior of $\max_i\frac{h_i}{g_i}$.  As a result, scheduling the SU with the smallest STPB channel gain for transmission does not change the sum-rate capacity scaling behavior up to a first order.  STSB channel gains only have a second order effect on the secondary network sum-rate. 

Furthermore, Theorem \ref{Theo: IL} reveals that the secondary network throughput scaling under $\CoSILFull$ and $\CoSILRed$ is controlled by a pre-log factor of $\frac{1}{\gamma_g}$ that is determined from the behavior of the CDF of the STPB channel gains around zero.  The pre-log factor is equal to $\frac{2}{c}$, $\frac{1}{m}$ and $1$ for the Weibull, Nakagami-$m$ and Rician-$K_f$ distributed STPB channel gains, respectively.  
The effect of $\gamma_g$ on $\RateILFull$ and $\RateILRed$ has an engineering interpretation. For a given fading model for the STPB channel gains, $\gamma_g$ is a measure for the proximity of the STPB channel power gains to zero.  That is, small values of $\gamma_g$ implies that the STPB channel gains take values close to zero with high probability, and vice versa. Thus, as $\gamma_g$ increases, the STPB channel gains become large, and consequently SUs reduce their transmission powers in order to meet the average interference power constraint at the PBS. As a result, the secondary network throughput decreases as $\gamma_g$ becomes large. 
 
In Appendix \ref{App: IL}, we show that the Lagrange multipliers $\mu_N$ converge to $\frac{1}{Q_{\rm ave}}$ in $\CoSILFull$.  This finding can be used to study the effect of $Q_{\rm ave}$ on the secondary network sum-rate under $\CoSILFull$.  Based on our analysis in Appendix \ref{App: IL}, we characterize the second order effects of $Q_{\rm ave}$ and fading parameters on the secondary network sum-rate under $\CoSILFull$ for finite number of SUs by bounding $\RateILFull$ from below and above as 
 \begin{eqnarray}\label{Eq: IL-up-low}
\lefteqn{\paren{1-\epsilon}\frac{1}{\gamma_g}\logp{N}+\logp{Q_{\rm ave}}+\frac{1}{\gamma_g}\logp{\eta_g \ES{h^{\gamma_g}}}+\BO{1} \leq \RateILFull\leq}\hspace{15cm}\nonumber\\
\lefteqn{ \paren{1+\epsilon}\frac{1}{\gamma_g}\logp{N}+\logp{Q_{\rm ave}}+\frac{1}{\gamma_g}\logp{\eta_g \ES{h^{\gamma_g}}}+\BO{1}}\hspace{9cm}
\end{eqnarray}  
for all $\epsilon>0$ and $N$ large enough ({\em i.e.,} see \eqref{Eq: FinalILFull}).  Hence, an increase in $Q_{\rm ave}$ leads to a logarithmic increase in $\RateILFull$, impliying that $Q_{\rm ave}$ has a logarithmic effect on $\RateILFull$.  Furthermore, the second order effects of the STSB and STPB fading models on $\RateILFull$ can be thought to be embodied in $\frac{1}{\gamma_g}\logp{\eta_g \ES{h^{\gamma_g}}}$, where $h$ is a generic nonnegative random variable with CDF $F_h(x)$. Since $\ES{h^{\gamma_g}}$ term depends on both STPB and STSB fading models, it is not possible to derive general insights about this term for arbitrary combinations of STSB-FMs and STPB-FMs.  Hence, we discuss the second order effects of the fading models on $\RateILFull$ when $\gamma_g$ is one, \emph{i.e.,} Rayleigh or Rician distributed STPB channel gains.  In this case, we have $\ES{h^{\gamma_g}}=1$, and $\eqref{Eq: IL-up-low}$ suggests that for a fixed number of SUs, $\RateILFull$ is predominantly affected by the parameters of the distribution of the STPB channel gains rather than those of the STSB channel gains. For Rician distributed STPB channel gains, the resulting second order term can be written as $\logp{\frac{K_f+1}{\e{K_f}}}$, which decreases with $K_f$. Note that larger $K_f$ implies more power in the line-of-sight component of the Rician fading, which, in turn, implies a larger interference power at the PBS. Hence, SUs decrease their transmission powers to meet the average interference power constraint at the PBS, which results in a reduction in the secondary network throughput.

\begin{remark}
Operating in the IL scenario does not necessary imply that the average transmit powers of SUs are infinite. It is easy to show that when the distribution of STPB channel gains belong to class $\mathcal{C}$-distributions with $\gamma_g>1$, the average transmit powers of SUs are finite. 
\end{remark}
 
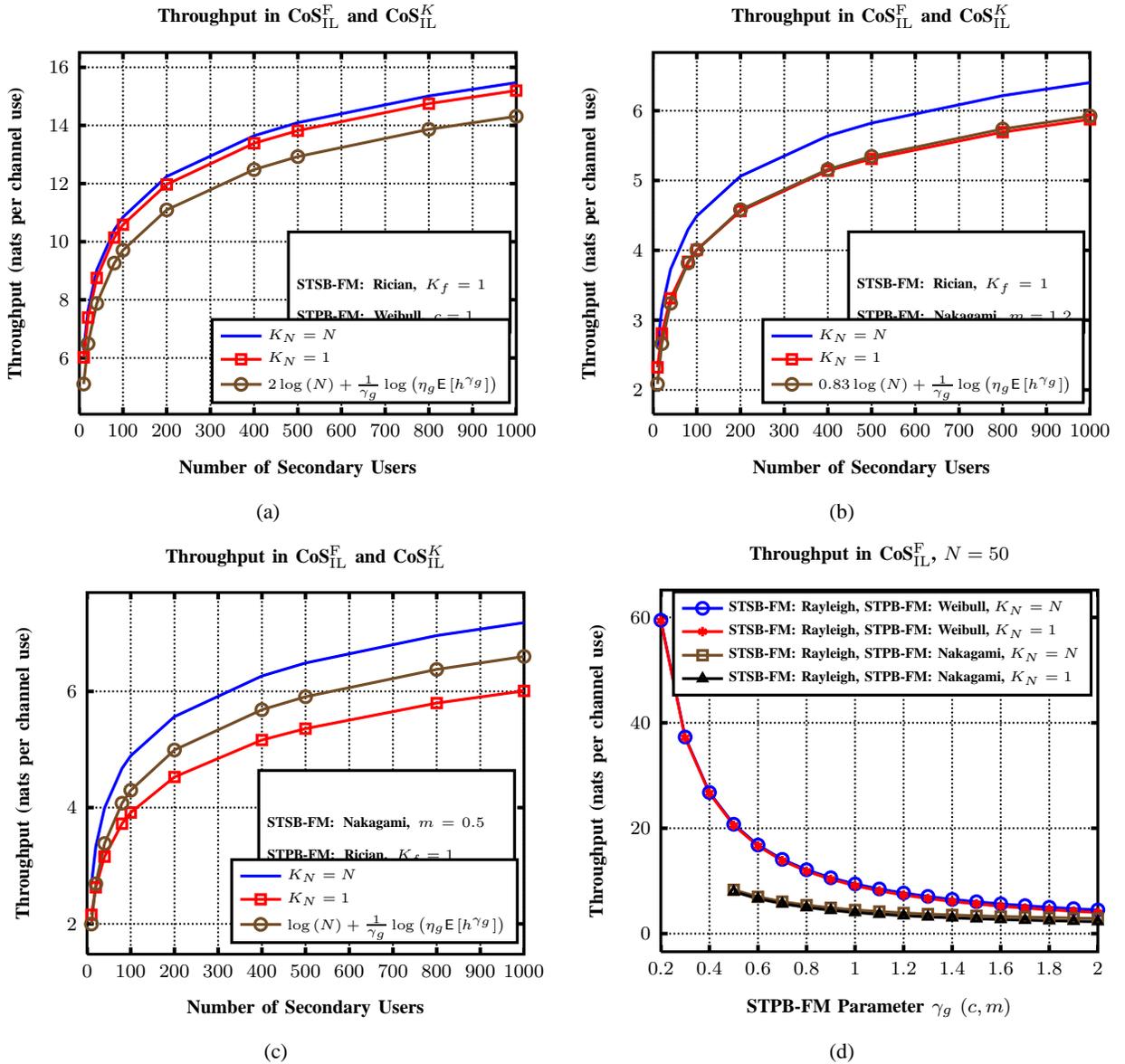
\begin{figure}[t]
\centering
\subfigure[]
{
\begin{tikzpicture}
\begin{axis}[title={Throughput in $\CoSILFull$ and $\CoSILRed$},legend style={yshift=-3.8cm, xshift=0.05cm},\axissetting]
                                      
\pfgsetting
\textbox{4.7cm}{1.5cm}{3cm}{Rician, $K_f=1$}{Weibull, $c=1$}
\addplot+[color=blue,mark=none] table [x=N,y=c1f]{RiWIL.dat};\addlegendentry{$\K=N$};
\addplot+[color=red,mark=square] table [x=N,y=c1k1]{RiWIL.dat};\addlegendentry{$\K=1$};
\addplot+[mark=o, mark size=2.5pt] table [x=N,y=2log]{RiWIL.dat};\addlegendentry{$2\logp{N}+\frac{1}{\gamma_g}\logp{\eta_g \ES{h^{\gamma_g}}}$};
\end{axis}
\end{tikzpicture}
\label{RiWIL}}
\subfigure[]
{
\begin{tikzpicture}
\begin{axis}[title={Throughput in $\CoSILFull$ and $\CoSILRed$},legend style={yshift=-3.8cm, xshift=0.05cm},\axissetting]
\pfgsetting
\textbox{4.6cm}{1.5cm}{3.2cm}{Rician, $K_f=1$}{Nakagami, $m=1.2$}
\addplot+[color=blue,mark=none] table [x=N,y=m1.2f]{RiNIL.dat};\addlegendentry{$\K=N$};
\addplot+[color=red,mark=square] table [x=N,y=m1.2k1]{RiNIL.dat};\addlegendentry{$\K=1$};
\addplot +[mark=o, mark size=2.5pt] table [x=N,y=0.8log]{RiNIL.dat};\addlegendentry{$0.83\logp{N}+\frac{1}{\gamma_g}\logp{\eta_g \ES{h^{\gamma_g}}}$};
\end{axis}
\end{tikzpicture}
\label{RiNIL}}
\subfigure[]
{
\begin{tikzpicture}
\begin{axis}[title={Throughput in $\CoSILFull$ and $\CoSILRed$},legend style={yshift=-3.8cm, xshift=0.05cm},\axissetting]
\pfgsetting
\textbox{4.4cm}{1.5cm}{3.5cm}{Nakagami, $m=0.5$}{Rician, $K_f=1$}
\addplot+[color=blue,mark=none] table [x=N,y=kf1f]{NRiIL.dat};\addlegendentry{$\K=N$};
\addplot+[color=red,mark=square] table [x=N,y=kf1k1]{NRiIL.dat};\addlegendentry{$\K=1$};
\addplot+[mark=o, mark size=2.5pt] table [x=N,y=log]{NRiIL.dat};\addlegendentry{$\logp{N}+\frac{1}{\gamma_g}\logp{\eta_g \ES{h^{\gamma_g}}}$};
\end{axis}
\end{tikzpicture}
\label{FNRiIL}}
\subfigure[]
{
\begin{tikzpicture}
\begin{axis}[title={Throughput in $\CoSILFull$, $N=50$ },legend style={yshift=0.1cm, xshift=0.05cm},xmin=0.2,xmax=2,xtick={0.2,0.4,0.6,0.8,1,1.2,1.4,1.6,1.8,2},grid=major, tick style={color=black, major tick length={0.10 cm}}, grid style={line width= 0.75pt, densely dotted, color= black}]
\pgfplotsset{
width= 8cm, 
every axis/.append style={line width=1.2pt},
label style={font=\bf\scriptsize}, 
ylabel style={yshift=-0.8em},
xlabel={STPB-FM Parameter $\gamma_g$ $(c, m)$},
ylabel={Throughput (nats per channel use)},
title style={font=\bf\scriptsize}, 
tick label style={font=\scriptsize,/pgf/number format/1000 sep={} },
tick style={ line width=1.5pt},
legend style={font=\bf\tiny,cells={anchor=west}}
}
\addplot+[color=blue,mark=o,mark size=2.5pt] table [x=N,y=WF]{ILvsParameter1.dat};\addlegendentry{\bf STSB-FM: Rayleigh, STPB-FM: Weibull, $\K=N$};
\addplot+[mark=asterisk] table [x=N,y=WR]{ILvsParameter1.dat};\addlegendentry{\bf STSB-FM: Rayleigh, STPB-FM: Weibull, $\K=1$};
\addplot+[mark=square] table [x=N1,y=NF]{ILvsParameter2.dat};\addlegendentry{\bf STSB-FM: Rayleigh, STPB-FM: Nakagami, $\K=N$};
\addplot+[mark=triangle] table [x=N1,y=NR]{ILvsParameter2.dat};\addlegendentry{\bf STSB-FM: Rayleigh, STPB-FM: Nakagami, $\K=1$};
\end{axis}
\end{tikzpicture}
\label{ILvsParameter}}
\caption{ Secondary network throughput scaling under $\CoSILFull$ ($\K=N$) and $\CoSILRed$ ($\K=1$) for different communication environments (a)-(c). Secondary network throughput under $\CoSTPILFull$ as a function of the STPB-FM parameter $\gamma_g$ for $N=50$ in different communication environments (d). $Q_{\rm ave}$ is set to 0dB.}
\label{IL}
\end{figure}

We plot the sum-rate scaling behavior of the secondary network under $\CoSILFull$ and $\CoSILRed$ as a function of the number of SUs for different STSB-FMs and STPB-FMs in Figs. 4(a)-(c). In these figures, the curves with $\K=N$ represent the secondary network sum-rate under $\CoSILFull$, and the curves with $\K=1$ represent the secondary network sum-rate under $\CoSILRed$.  $Q_{\rm ave}$ is set to 0dB. Similar qualitative behavior continues to hold for other values of $Q_{\rm ave}$. In Fig. \ref{RiWIL}, STSB channel gains are distributed according to the Rician fading model with $K_f=1$, and STPB channel gains are distributed according to the Weibull fading model with $c=1$. 

Fig. \ref{RiWIL} shows that the secondary network sum-rate scales according to $\frac{2}{c}\logp{N}$ when STPB channel gains are Weibull distributed; a behavior which was predicted by Theorem \ref{Theo: IL}.  In Fig. \ref{RiNIL}, STSB channel gains are distributed according to the Rician fading model with $K_f=1$, and STPB channel gains are distributed according to the Nakagami-$m$ fading model with $m=1.2$.  Fig. \ref{RiNIL} reveals that the secondary network sum-rate scales according to $\frac{1}{m}\logp{N}$ when STPB channel gains are Nakagami-$m$ distributed, which is in accordance with Theorem \ref{Theo: IL}.  In Fig. \ref{FNRiIL}, STSB channel gains are distributed according to the Nakagami-$m$ fading model with $m=0.5$, and STPB channel gains are distributed according to the Rician fading model with $K_f = 1$.  As Fig. \ref{FNRiIL} shows, the secondary network sum-rate scales according to $\logp{N}$ when STPB channel gains are distributed according to the Rician fading model as predicted by Theorem \ref{Theo: IL}.  Fig. 4(a)-(c) demonstrate that the sum-rate loss due to implementing the $K$-SCG feedback protocol is within one nats per channel use when compared to the full-feedback protocol, which signifies that the $K$-SCG feedback protocol is an effective primary-secondary feedback reduction policy for interference limited cognitive radio networks.

Fig. \ref{ILvsParameter} depicts the dependence of the secondary network sum-rate under $\CoSILFull$ and $\CoSILRed$ on $\gamma_g$ in a CMAC with $N=50$.  In this figure, the curves with $\K=N$ represent the secondary network throughput under $\CoSILFull$, and the curves with $\K=1$ represent the secondary network throughput under $\CoSILRed$.  STSB channel gains are Rayleigh distributed, and STPB channel gains are Weibull and Nakagami-$m$ distributed.  In Fig. \ref{ILvsParameter}, as $\gamma_g$  increases, STPB channel gains become large, and SUs reduce their transmission powers to meet the interference constraint. Thus, the secondary network throughput drops as discussed above.


The next theorem establishes the secondary network sum-rate scaling behavior under $\CoSIPILFull$ and $\CoSIPILRed$.  In Appendix \ref{App: IPIL}, we give a detailed proof for this theorem.
 \begin{theorem}\label{Theo: IPIL}
Let $\RateIPILFull$ and $\RateIPILRed$ be the secondary network throughput under $\CoSIPILFull$ and $\CoSIPILRed$ for $0<\K\leq N$, respectively. Then, 
\begin{eqnarray}
\lim_{N\ra\infty}\frac{\RateIPILRed}{\logp{N}}=\lim_{N\ra\infty}\frac{\RateIPILFull}{\logp{N}}=\min\paren{1,\frac{1}{\gamma_g}}.\nonumber
\end{eqnarray}
\end{theorem}
\begin{IEEEproof}
Please see Appendix \ref{App: IPIL}.
\end{IEEEproof}

Theorem \ref{Theo: IPIL} establishes the logarithmic scaling behavior of  the secondary network sum-rate under $\CoSIPILFull$ and $\CoSIPILRed$ as a function of the number of SUs when the CDFs of STSB and STPB channel gains belong to class $\mathcal{C}$-distributions.  For $\CoSIPILRed$, the scaling behavior does not depend on the number of STPB channel gains available at the SBS. Hence, even for $\K=\BO{1}$, a secondary network under $\CoSIPILRed$ can achieve the same scaling behavior as the one achieved under $\CoSIPILFull$, which implies a tremendous reduction in the primary-secondary feedback load. 

Theorem \ref{Theo: IPIL} also reveals the effect of parameters of the STPB fading model on the scaling behavior of $\RateIPILFull$ and $\RateIPILRed$, which appears as the pre-log factor of $\min\paren{1,\frac{1}{\gamma_g}}$. This effect has the following interpretation. For $\gamma_g<1$, random STPB channel gains take values close to zero with high probability.  As a result, the average interference power constraint becomes increasingly looser, and the transmission powers of SUs become mainly limited by the individual average power constraints, \emph{i.e.,} secondary network behaves as a primary MAC with individual power constraints only.  In Lemma \ref{Lem: Primary-MAC} in Appendix \ref{App: IPIL}, we show that the sum-rate of a primary MAC  with individual power constraints $\RateIPM$ scales according to $\log{N}$.  Hence, the secondary network throughput scales according to $\logp{N}$ for $\gamma_g<1$.  For $\gamma_g>1$, random STPB channel gains take large values away from zero with high probability, when compared with the case of $\gamma_g<1$.  Thus, the average interference power constraint becomes more stringent, and the secondary network behaves as an IL network. This leads to the result that the secondary network throughput scales according to $\frac{1}{\gamma_g}\logp{N}$ for $\gamma_g>1$.   

From a more heuristic perspective, the effect of the pre-log factor $\min\paren{1,\frac{1}{\gamma_g}}$ has the following interpretation.  By removing the interference power constraint from an IPIL network, we obtain a primary MAC with individual power constraints whose sum-rate $\RateIPM$ can be shown to scale according to $\logp{N}$.  On the other hand, by removing the individual power constraints from an IPIL network, we obtain an IL network whose throughput $\RateILFull$ can be shown to scale according to $\frac{1}{\gamma_g}\logp{N}$.  Thus, $\RateIPILFull$ is upper bounded by both $\RateILFull$ and $\RateIPM$. Depending on the value of $\gamma_g$, one of the upper bounds bites.  That is, for $\gamma_g<1$, $\RateIPM$ bound is tighter than $\RateILFull$, and as a result the secondary network throughput scales according to $\logp{N}$. For $\gamma_g>1$, $\RateILFull$ bound is tighter than $\RateIPM$, and the secondary network throughput scales according to $\frac{1}{\gamma_g}\logp{N}$.  It should be noted these arguments can only provide us with an upper bound. More analysis is needed to establish the lower bounds with the same scaling behavior, {\em i.e.,} see Appendix \ref{App: IPIL}.

\begin{figure}[t]
\centering
\subfigure[]
{
\begin{tikzpicture}
\begin{axis}[title={Throughput in $\CoSILFull$ and $\CoSILRed$},legend style={yshift=-3.8cm, xshift=0.05cm},\axissetting]
                                      
\pfgsetting
\textbox{4.75cm}{1.5cm}{2.85cm}{ Rayleigh }{Weibull, $c=1.5$}
\addplot+[color=blue,mark=none] table [x=N,y=log]{RNIPIL.dat};\addlegendentry{$\logp{N}+\logp{P_{\rm ave}}$};
\addplot+[color=red,mark=square] table [x=N,y=fc1.5]{RNIPIL.dat};\addlegendentry{$\K=N$};
\addplot+[mark=o, mark size=2.5pt] table [x=N,y=k1c1.5]{RNIPIL.dat};\addlegendentry{$\K=1$};
\end{axis}
\end{tikzpicture}
\label{RNIPIL1}}
\subfigure[]
{
\begin{tikzpicture}
\begin{axis}[title={Throughput in $\CoSILFull$ and $\CoSILRed$},legend style={yshift=-3.8cm, xshift=0.05cm},\axissetting]
\pfgsetting
\textbox{4.75cm}{1.5cm}{2.85cm}{ Rayleigh }{Weibull, $c=2.5$}
\addplot+[color=blue,mark=none] table [x=N,y=fc2.5]{RNIPIL.dat};\addlegendentry{$\K=N$};
\addplot+[color=red,mark=square] table [x=N,y=0.8log ]{RNIPIL.dat};\addlegendentry{$0.8\logp{N}+\frac{1}{\gamma_g}\logp{\eta_g \ES{h^{\gamma_g}}}$};
\addplot +[mark=o, mark size=2.5pt] table [x=N,y=k1c2.5]{RNIPIL.dat};\addlegendentry{$\K=1$};
\end{axis}
\end{tikzpicture}
\label{RNIPIL2}}
\caption{ Secondary network throughput scaling under $\CoSIPILFull$ ($\K=N$) and $\CoSIPILRed$ ($\K=1$) for different communication environments. $P_{\rm ave}$ and $Q_{\rm ave}$ are set to 15dB and 0dB, respectively.}
\label{IPIL}
\end{figure}
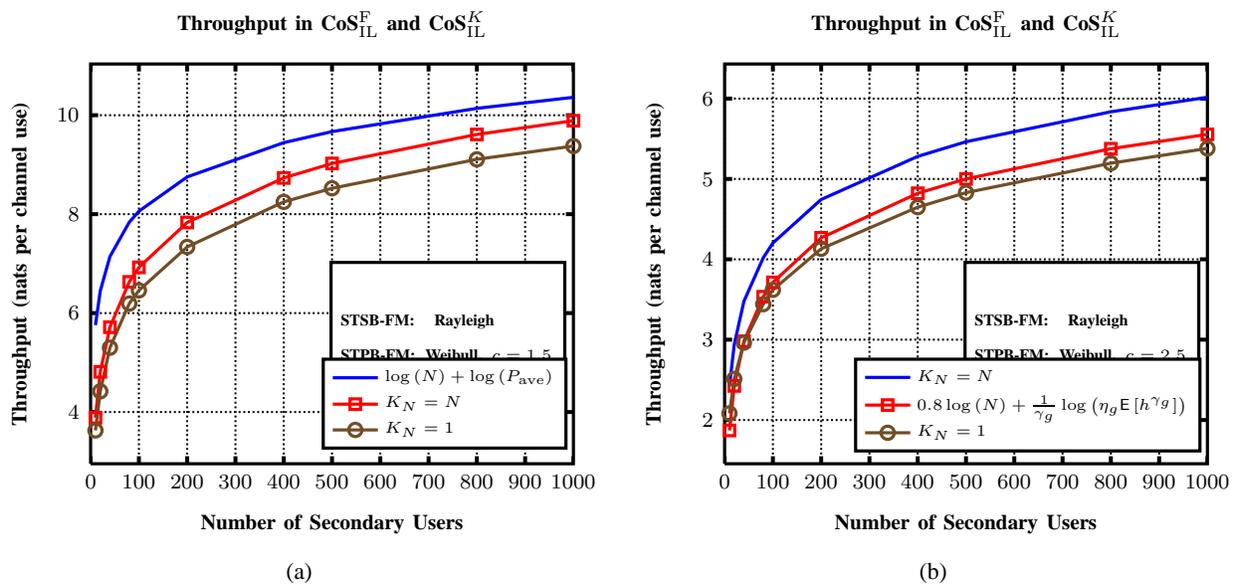

We demonstrate the sum-rate scaling behavior of the secondary network under $\CoSIPILFull$ and $\CoSIPILRed$ as a function of the number of SUs in Fig. \ref{IPIL}.  In this figure, the curves with $\K=N$ represent the secondary network sum-rate under $\CoSIPILFull$, and the curves with $\K=1$ represent the secondary network sum-rate under $\CoSIPILRed$.  $P_{\rm ave}$ and $Q_{\rm ave}$ are set to 15dB and 0dB, respectively.  Similar qualitative behavior continues to hold for other values of $P_{\rm ave}$ and $Q_{\rm ave}$. Fig. \ref{RNIPIL1} illustrates the secondary network throughput scaling when STSB channel gains are distributed according to the Rayleigh fading model, and STPB channel gains are distributed according to the Weibull fading model with $c=1.5$.  Fig. \ref{RNIPIL1} indicates that the throughput of the secondary network scales according to $\logp{N}$ for $c\leq 2$ as predicted by Theorem \ref{Theo: IL}.  The $\logp{N}+\logp{P_{\rm ave}}$ curve represents the scaling behavior of the primary multiple access channel with individual power constraints obtained by removing the interference power constraint from the original IPIL network.  Closeness of this curve to our simulated data rates confirms that an IPIL network behaves similar to a primary MAC with individual power constraints for $\gamma_g<1$. 

Fig. \ref{RNIPIL2} represents the secondary network sum-rate when STSB channel gains are distributed according to the Rayleigh fading model, and STPB channel gains are distributed according to the Weibull fading model with $c=2.5$.  As this figure shows, the secondary network throughput scales according to $\frac{2}{c}\logp{N}$ in accordance with the MDGs predicted by Theorem \ref{IL}.  In Fig. \ref{RNIPIL2}, the $0.8\logp{N}+\frac{1}{\gamma_g}\logp{\eta_g \ES{h^{\gamma_g}}}$ curve quantifies the sum-rate scaling behavior of the IL network obtained by removing the individual power constraints from the original IPIL network.  Closeness of this curve to our simulated data rates confirms that an IPIL network behaves similar to an IL network for $\gamma_g>1$.  Finally, Figs. 4(a)-(b) show that the throughput loss arising from implementing the $K$-SCG feedback protocol is within one nats per channel use again when compared to the full-feedback protocol, which signifies that the $K$-SCG feedback protocol is an effective primary-secondary feedback reduction policy in this case, too. 

\section{conclusion}\label{sec: conclusion}
 In this paper, we have analyzed the secondary network sum-rate scaling behavior for cognitive radio multiple access channels in feedback limited communication scenarios.  To this end, we have first introduced an efficient primary-secondary feedback protocol called the $K$-smallest channel gains ($K$-SCGs) feedback protocol in which the PBS feeds back the $\K$ smallest STPB channel gains to the SBS (out of $N$ STPB channel gains).  The effect of the $K$-SCG feedback protocol on the secondary network sum-rate scaling behavior has been studied for three different network types when the transmission powers of secondary users (SUs) are optimally allocated. The network types considered are the total-power-and-interference-limited (TPIL), interference-limited (IL) and individual-power-and-interference-limited (IPIL) networks.  In TPIL networks, transmit powers of SUs are limited by an average total power constraint as well as a constraint on the average total interference power that they cause to the PBS. On the other hand, transmit powers of SUs are limited by a constraint only on the average total interference power at the PBS for IL networks. In the case of IPIL networks, transmit powers of SUs are limited by individual average power constraints as well as a constraint on the average total interference power at the PBS. 
 
For each network type considered, we have derived a sufficient condition on $\K$ such that the $K$-SCG feedback protocol is asymptotically optimal. It has been shown that for $\K=N^\delta$ with $\delta\in\left(0,1\right)$, the $K$-SCG feedback protocol is asymptotically optimal in TPIL networks, \emph{i.e.,} the secondary network sum-rate under $K$-SCG and full-feedback protocols scales according to $\frac{1}{n_h}\log\logp{N}$, where $n_h$ is a parameter obtained from the distribution of the STSB channel power gains and $N$ is the number of SUs. In TPIL networks, it has also been shown that for $\K$ growing to infinity at a rate $\K=\LO{N}$, the interference power at the PBS converges to zero almost surely and in mean as $N$ becomes large.  Once this happens, the secondary network just requires the indices of SUs having the $\K$ smallest STSB channel gains for the jointly optimal power allocation and scheduling as $N$ becomes large.  For IL networks, It has been shown that having $\K=\BO{1}$ is enough for the asymptotical optimality of the $K$-SCG feedback protocol.  In this case, the secondary network sum-rate under $K$-SCG and full-feedback protocols scales according to $\frac{1}{\gamma_g}\logp{N}$ , where $\gamma_g$ is a parameter controlling the decay rate of the CDF of the STPB channel gains around zero. $\K=\BO{1}$ is also enough for the asymptotical optimality of the $K$-SCG for IPIL networks.  In this case, the secondary network sum-rate under $K$-SCG and full-feedback protocols scales according to $\min\paren{1,\frac{1}{\gamma_g}}\logp{N}$.  

\appendices
\section{Asymptotic Behavior of Extreme Order Statistics}\label{app: concentration}
In this appendix, we study the concentration behavior of the extreme order statistic of a sequence of i.i.d. random variables as the number of elements in the sequence grows large. Later, this result will play a central role in deriving the cognitive radio throughput scaling behavior in different communication scenarios. To this end, let $\left\{Y_i\right\}_{i=1}^N$ be a sequence of i.i.d random variables with a common probability distribution function $F(x)$.  We assume that $\lim_{x\ra \infty} F(x)=1$, $F(x)<1$ for $x<\infty$, and there exists $x_0<\infty$ such that $F\paren{x_1}< F\paren{x_2}$ whenever $x_0<x_1<x_2<\infty$.  We call a CDF possessing these properties an \emph{eventually increasing} CDF. Let $G(x)$ be a function such that $\label{Eq: F-asymp} \lim_{x\ra\infty}G(x)\paren{1-F(x)}=1$.  We say $G(x)$ characterizes the tail behavior of $F(x)$.  Without loss of generality, we assume that $G(x): \left(C,\infty\right) \mapsto \R_+$, and $G(x)$ is strictly increasing on $\left(C,\infty\right)$.  Note that $G(x)$ is invertible, and we denote its functional inverse as $G^{-1}(x)$.\footnotemark[1]\footnotetext[1]{Since $F(x)$ is eventually increasing, $G(x)$ is also eventually increasing and tends to infinity as $x$ tends to infinity.  Thus, we can find a large positive constant $C$ such that $G(x)$ is strictly increasing on $\left(C,\infty\right)$, and $G(x)$ is invertible on this interval.}  Let $Y^\star_N$ be the extreme order statistic of $\left\{Y_i\right\}_{i=1}^N$, {\em i.e.,} $Y^\star_N=\max\limits_{1\leq i\leq N}Y_i$.  Let $F^\star_N(x)$ be the CDF of $Y^\star_N$. 

The next lemma establishes an important concentration property for $Y^\star_N$.  This result will be used to study the convergence behavior of $Y^\star_N$.  It also shows that the asymptotic behavior of $Y^\star_N$ is characterized by $G(x)$ as $N$ tends to infinity. Note that the class of eventually increasing distributions covers the class $\cal{C}$ distributions.
\begin{lemma}\label{Lem: Concentration}
Let $\brparen{Y_i}_{i=1}^N$ be a sequence of i.i.d. random variables with an eventually increasing common CDF $F(x)$ whose tail behavior is characterized by $G(x)$.  Also, let $Y^\star_N=\max_{1\leq i\leq N}Y_i$.  Then, for any $\epsilon$ belonging to $(0, 1)$, we have 
\begin{eqnarray}
\lim_{N \ra \infty} \PR{\inv{G}{N^{1-\epsilon}}< Y^\star_N \leq \inv{G}{N^{1+\epsilon}}} = 1.
\end{eqnarray}
\end{lemma}
\begin{IEEEproof}
For $x$ large enough, we can express $F_N^\star(x)$ as $F_N^\star(x) = \e{-N\BT{\frac{-1}{G(x)}}}$ since $\lim_{x\ra\infty}G(x)\paren{1-F(x)}=1$.  By using this expression for $F_N^\star(x)$, we have:
\begin{eqnarray}\label{Eq: Up-bound}
\PR{Y^\star_N\leq \inv{G}{N^{1+\epsilon }}}&=&\e{-N\BT{\frac{1}{N^{1+\epsilon }}}}\nonumber\\
&=&1-\BT{\frac{1}{N^{\epsilon} }},
\end{eqnarray}
and
\begin{eqnarray}\label{Eq: Low-bound}
\PR{Y^\star_N\leq \inv{G}{N^{1-\epsilon }}}&=&\e{-N\BT{\frac{1}{N^{1-\epsilon }}}}\nonumber\\
&=&\e{-\BT{N^{\epsilon} }}.
\end{eqnarray}
for all $\epsilon \in (0, 1)$ and $N$ large enough.  Therefore, we have
\begin{eqnarray}\label{Eq: prob-Conv}
\PR{\inv{G}{N^{1-\epsilon }}< Y^\star_N \leq \inv{G}{N^{1+\epsilon }}}&=&1-\BT{\frac{1}{N^{\epsilon} }},
\end{eqnarray}
which implies that $Y^\star_N$ lies in $\left[\inv{G}{N^{1-\epsilon }}, \inv{G}{N^{1+\epsilon }}\right]$ with probability approaching 1 as $N$ grows large. 
\end{IEEEproof}
\section{Throughput Scaling in Total-Power-And-Interference-Limited Scenario}\label{App: TPIL}
In this appendix, without loss of generality, we establish the throughput scaling behaviour of TPIL networks when the interference plus noise power is equal to 1. The asymptotic behavior of $\RateTPILFull$ and $\RateTPILRed$ depends on the asymptotic behavior of $\X{\lambda_N}{\mu_N}=\max\limits_{1\leq i\leq N }\frac{h_i}{\lambda_N+\mu_Ng_i}$ and $\Xr{\lambda_N}{\mu_N}=\max_{1\leq i\leq \K }\frac{h_{\pi(i)}}{\lambda_N+\mu_Ng_{\pi(i)}}$ as a function of $N$, respectively.  Since Lagrange multipliers $\lambda_N$ and $\mu_N$, which can be different for different communication scenarios, vary with $N$, $\left\{\frac{h_i}{\lambda_N+\mu_Ng_i}\right\}_{i=1}^N$ and $\left\{\frac{h_{\pi(i)}}{\lambda_N+\mu_N g_{\pi(i)}}\right\}_{i=1}^{\K}$ form triangular arrays of random variables. This complicates the analysis to some extent.  Hence, to simplify our analysis, we start with the characterization of the asymptotic behavior of $\X{\lambda}{\mu}$ and $\Xr{\lambda}{\mu}$ for some fixed non-negative real numbers $\lambda$ and $\mu$.  Then, we use this result to obtain the asymptotic behavior of $\RateTPILFull$ and $\RateTPILRed$ for large values of $N$. 

The next three assisting lemmas play a key role in the proof Theorem \ref{Theo: TPIL}. Recall that all distribution functions belong to the class $\mathcal{C}$ distributions, and therefore, the various parameters such as $\alpha, l, \beta, n, H(x), \gamma$ and $\eta$ appearing in our analysis below are as defined in Definition \ref{Def1}. For specific channel models, they are given in Table \ref{Table: Fading Parameters}. 

Let $\tilde{R}\paren{N, \lambda, \mu} = \ES{\logp{\X{\lambda}{\mu}}\I{\X{\lambda}{\mu}\geq1}}$.  In the next lemma, we establish the asymptotic behavior of $\tilde{R}\paren{N, \lambda, 0}$, which will be helpful in upper bounding the sum-rate in $\CoSTPILFull$.     
\begin{lemma}\label{Lem: Conv-Xt}
For $\lambda > 0$, $\lim_{N\ra\infty}\frac{\tilde{R}\paren{N,\lambda,0}}{\logp{\frac{1}{\lambda}\paren{\frac{1}{\beta_h}\logp{\alpha_h N}}^\frac{1}{n_h}}}=1$. 
\end{lemma}
\begin{IEEEproof}
Let $\Xt{\lambda}{0}=\frac{\logp{\X{\lambda}{0}}}{\logp{\frac{1}{\lambda}\paren{\frac{1}{\beta_h}\logp{\alpha_h N}}^\frac{1}{n_h}}}\I{\X{\lambda}{0}\geq 1}$.  Hence, to show the desired result, it is enough to show $\lim_{N\ra\infty}\ES{\Xt{\lambda}{0}}=1$. Note that $\X{\lambda}{0}=\frac{1}{\lambda}\h{N}$, where $\h{N}=\max_{1\leq i \leq N}h_i$.  Recall $F_h(x)$ is the probability distribution function common to all $h_i$, $1 \leq i \leq N$.  From \eqref{Eq: tail-condition-1}, the tail behavior of $F_h(x)$ is characterized as $G(x)=\frac{x^{-l_h}}{\alpha_h}\e{\beta_h x^{n_h}-H(x)}$.  Hence, we can write $\inv{G}{x}=\paren{\frac{\frac{1}{\beta_h}\logp{\alpha_h x}}{1-\frac{l_h\logp{\inv{G}{x}}+H\paren{\inv{G}{x}}}{\beta\paren{\inv{G}{x}}^{n_h}}}}^\frac{1}{n_h}$.  Note that $1-\frac{l_h\logp{\inv{G}{x}}+H\paren{\inv{G}{x}}}{\beta_h\paren{\inv{G}{x}}^{n_h}}=1-\LO{1}$.  Thus, $\inv{G}{x}=\paren{\frac{1}{\beta_h}\logp{\alpha_h x}}^\frac{1}{n_h}\paren{1+\LO{1}}$.  Using Lemma \ref{Lem: Concentration}, we have 
\begin{eqnarray}\label{Eq: hstar-concentration}
\lefteqn{\PR{\inv{G}{\alpha_h^{-\epsilon}N^{1-\epsilon}}<\h{N}\leq \inv{G}{\alpha_h^{\epsilon}N^{1+\epsilon}}}}\hspace{17cm}\nonumber\\
\lefteqn{=\PR{\paren{\frac{1}{\beta_h}\logp{\alpha_h N}}^\frac{1}{n_h}\paren{1-\epsilon}^\frac{1}{n}\paren{1+\LO{1}} <\h{N}\leq \paren{\frac{1}{\beta_h}\logp{\alpha_h N}}^\frac{1}{n_h}\paren{1+\epsilon}^\frac{1}{n}\paren{1+\LO{1}}}}\hspace{15cm}\nonumber\\
\lefteqn{=1-\BT{\frac{1}{N^{\epsilon}}}}\hspace{15cm}
\end{eqnarray}
for $N$ large enough.  This final result implies that $\Xt{\lambda}{0} \xrightarrow{i.p.}1$ as $N$ goes to infinity, where $i.p.$ stands for in probability.  

Since convergence in probability does not alway imply convergence in mean, we need to prove that $\left\{\Xt{\lambda}{0}\right\}_{N=1}^\infty$ is a uniformly integrable collection of random variables, or equivalently $\lim_{C \ra\infty}\sup_{N \geq 1} \ES{\Xt{\lambda}{0}\I{\Xt{\lambda}{0}>C}}=0$, to conclude the proof \cite{Billingsley}. To prove uniform integrability of $\left\{\Xt{\lambda}{0}\right\}_{N=1}^\infty$, it is enough to prove that the collection of random variables $\brparen{\Xt{\lambda}{0}}_{N \geq N_0}$ is uniformly integrable for some finite positive integer $N_0$ \cite{Gut}.  To this end, we will show that for any given $\epsilon_1$, $\epsilon_2$ and $\epsilon_3$ satisfying $\epsilon_1>0$, $0<\epsilon_2<\beta_h$ and $\epsilon_3>0$, we can find large enough positive constants $N_0$ and $C_0$ such that 
\begin{eqnarray}\label{Eq: sup-upperbound}
\sup_{N\geq N_0}\ES{\Xt{\lambda}{0}\I{\Xt{\lambda}{0}>C}}
&\leq&\paren{1+\epsilon_1}\paren{1+\epsilon_3}\alpha N_0\lambda^l\frac{C B_{N_0}^{lC}}{\e{\paren{\beta-\epsilon_2} \lambda^nB_{N0}^{nC}}}\sum_{i=0}^{\infty}\frac{\paren{i+2}B_{N_0}^{iCl}}{\e{ B_{N0}^{niC}}}
\end{eqnarray}
for all $C$ greater than $C_0$, where $B_N=\frac{1}{\lambda}\paren{\frac{1}{\beta_h}\logp{\alpha_h N}}^\frac{1}{n_h}$.  Observe that $\sum_{i=0}^{\infty}\frac{\paren{i+2}B_{N_0}^{iCl}}{\e{ B_{N0}^{niC}}}$ is a convergent series for $N_0$ large enough.  Thus, $\lim\limits_{C\ra\infty}\sup\limits_{N\geq N_0}\ES{\Xt{\lambda}{0}\I{\Xt{\lambda}{0}>C}}=0$, which implies $\left\{\Xt{\lambda}{0}\right\}_{N=1}^{\infty}$ is uniformly integrable.  The rest of the proof is devoted to show that \eqref{Eq: sup-upperbound} holds.  

For $C>0$, $\ES{\Xt{\lambda}{0}\I{\Xt{\lambda}{0}>C}}$ can be upper bounded as 
\begin{eqnarray}\label{Eq: uni-integ-UB}
\ES{\Xt{\lambda}{0}\I{\Xt{\lambda}{0}>C}} &\leq&\sum_{i=1}^{\infty}\paren{i+1}C \PR{iC< \Xt{\lambda}{0}\leq\paren{i+1}C}\nonumber\\
&\leq&\sum_{i=1}^{\infty}\paren{i+1}C\PR{\lambda B_N^{iC}< \h{N}}.
\end{eqnarray}
For any given $\epsilon_1>0$, $<\epsilon_2<\beta_h$, $\epsilon_3>0$ and $N$ large enough, $\PR{\lambda B_N^{iC}< \h{N}}$ can be upper bounded as
\begin{eqnarray}
\PR{\lambda B_N^{iC}<\h{N}}&=&1-F_h^N\paren{\lambda B_N^{iC}}\nonumber\\
&\stackrel{\rm (a)}{\leq} &1-\e{N\logp{1-\frac{\paren{1+\epsilon_1}\alpha_h \paren{\lambda B_N^{iC}}^{l_h}}{\e{\beta_h \paren{\lambda B_N^{iC}}^{n_h}-H\paren{\lambda B_N^{iC}}}}}}\nonumber\\
&\stackrel{\rm (b)}{\leq} &1-\e{N\logp{1-\frac{\paren{1+\epsilon_1}\alpha_h \paren{\lambda B_N^{iC}}^{l_h}}{\e{\paren{\beta_h-\epsilon_2} \paren{\lambda B_N^{iC}}^{n_h}}}}}\nonumber\\
&\stackrel{\rm (c)}{\leq}& 1-\e{-\paren{1+\epsilon_1}\paren{1+\epsilon_3}\frac{\alpha_h N\paren{\lambda B_N^{iC}}^{l_h}}{\e{\paren{\beta_h-\epsilon_2} \paren{\lambda B_N^{iC}}^{n_h}}}}\nonumber\\
&\stackrel{\rm (d)}{\leq}& \paren{1+\epsilon_1}\paren{1+\epsilon_3}\frac{\alpha_h N\paren{\lambda B_N^{iC}}^{l_h}}{\e{\paren{\beta_h-\epsilon_2} \paren{\lambda B_N^{iC}}^{n_h}}},\nonumber
\end{eqnarray}
where (a) follows from the fact that for all $\epsilon_1>0$ and $x$ large enough, $F_h(x)$ can be lower bounded as $F_h(x)\geq 1-\frac{\paren{1+\epsilon_1}\alpha x^{l_h}}{\e{\beta_h x^{n_h}-H\paren{x}}}$ by Definition \ref{Def1}, (b) follows from the fact that for all $\epsilon_2 \in (0, \beta)$ and $x$ large enough, $H\paren{x}$ can be upper bounded as $H\paren{x}<\epsilon_2x^{n_h}$ since $H\paren{x} = \LO{x^{n_h}}$, (c) follows from the fact that for all $\epsilon_3>0$ and $x>0$ close enough to zero, $\logp{1-x}$ can be lower bounded as $\logp{1-x}\geq-\paren{1+\epsilon_3}x$ since $\lim_{x\downarrow0}\frac{\logp{1-x}}{x}=-1$, and finally (d) follows form the fact that $\e{-x}\geq 1-x$ for $x\geq 0$.  Thus, for all $\epsilon_1>0$, $<\epsilon_2<\beta_h$, $\epsilon_3>0$ and $N$ large enough, $\ES{\Xt{\lambda}{0}\I{\Xt{\lambda}{0}>C}}$ can be upper bounded as 
\begin{eqnarray}
\ES{\Xt{\lambda}{0}\I{\Xt{\lambda}{0}>C}}&\leq&\sum_{i=1}^{\infty}\paren{i+1}C\paren{1+\epsilon_1}\paren{1+\epsilon_3}\frac{\alpha_h N\paren{\lambda B_N^{iC}}^{l_h}}{\e{\paren{\beta_h-\epsilon_2} \paren{\lambda B_N^{iC}}^{n_h}}}. \nonumber
\end{eqnarray}

Now, we show that there exist large enough positive constants $N_0$ and $C_0$ such that $\frac{ N\paren{\lambda B_N^{iC}}^{l_h}}{\e{\paren{\beta_h-\epsilon_2}\paren{\lambda B_N^{iC}}^{n_h}}}$ is a decreasing function of $N$ for $N\geq N_0$ whenever $C$ is fixed but larger than $C_0$.  By considering $N$ as a positive real number with a slight abuse of notation, the first derivative of $\frac{ N\paren{\lambda B_N^{iC}}^{l_h}}{\e{\paren{\beta_h-\epsilon_2}\paren{\lambda B_N^{iC}}^{n_h}}}$ with respect to $N$ can be obtained as
\begin{eqnarray}
\frac{\partial}{\partial N}\frac{ N\paren{\lambda B_N^{iC}}^{l_n}}{\e{\paren{\beta_h-\epsilon_2} \paren{\lambda B_N^{iC}}^{n_h}}}&=&\frac{iCB_N^{(iC-1)n_h}\paren{\lambda B_N^{iC}}^{l_h}}{\e{\paren{\beta_h-\epsilon_2} \paren{\lambda B_N^{iC}}^{n_h}}}\paren{\frac{1}{iCB_N^{(iC-1)n_h}}+\frac{l}{n_h\lambda^{n_h}\beta_h B_N^{iCn_h} }-\frac{\paren{\beta_h-\epsilon_2}}{\beta_h }}. \nonumber
\end{eqnarray}
This final equation implies that we can find large enough positive constants $C_0$ and $N_0$ such that $B_{N_0}>1$ and $\frac{1}{CB_{N}^{(iC-1)n_h}}+\frac{l}{n_h\lambda^{n_h}\beta B_{N}^{iCn_h} }-\frac{\paren{\beta_h-\epsilon_2}}{\beta_h}< 0$ for $C\geq C_0$, $N\geq N_0$ and $i\in \N$.
Finally, we have
\begin{eqnarray}
\sup_{N\geq N_0}\ES{\Xt{\lambda}{0}\I{\Xt{\lambda}{0}>C}}&\leq&\sup_{N\geq N_0}\sum_{i=1}^{\infty}\paren{i+1}C\paren{1+\epsilon_1}\paren{1+\epsilon_3}\frac{\alpha_h N\paren{\lambda B_N^{iC}}^{l_h}}{\e{\paren{\beta_h-\epsilon_2} \paren{\lambda B_N^{iC}}^{n_h}}}\nonumber\\
&\stackrel{\rm (a)}{=}&\sum_{i=1}^{\infty}\paren{i+1}C\paren{1+\epsilon_1}\paren{1+\epsilon_3}\frac{\alpha_h N_0\paren{\lambda B_{N_0}^{iC}}^{l_h}}{\e{\paren{\beta_h-\epsilon_2} \paren{\lambda B_{N0}^{iC}}^{n_h}}}\nonumber \hspace{2.5cm}
\end{eqnarray}
\begin{eqnarray}
\hspace{4.7cm} &=&C\paren{1+\epsilon_1}\paren{1+\epsilon_3}\alpha_h N_0\lambda^{l_h} B_{N_0}^{Cl_h}\sum_{i=0}^{\infty}\frac{\paren{i+2}B_{N_0}^{iCl_h}}{\e{\paren{\beta_h-\epsilon_2} \lambda^{n_h} B_{N0}^{iCn_h}B_{N0}^{Cn_h}}}\nonumber\\
&\stackrel{\rm (b)}{\leq}&\paren{1+\epsilon_1}\paren{1+\epsilon_3}\alpha_h N_0\lambda^{l_h}\frac{C B_{N_0}^{Cl_h}}{\e{\paren{\beta_h-\epsilon_2} \lambda^{n_h}B_{N0}^{Cn_h}}}\sum_{i=0}^{\infty}\frac{\paren{i+2}B_{N_0}^{iCl_h}}{\e{ B_{N0}^{iCn_h}}},
\end{eqnarray}
where (a) follows from the fact that $\frac{ N\paren{\lambda B_N^{iC}}^{l_h}}{\e{\paren{\beta_h-\epsilon_2}\paren{\lambda B_N^{iC}}^{n_h}}}$ is decreasing with $N$ for $N\geq N_0$ and $C \geq C_0$, and (b) follows from the fact that $B_{N0}^{iCn_h}B_{N0}^{Cn_h}\geq B_{N0}^{Cn_h}+B_{N0}^{iCn_h}$ for $N_0$ large enough, which completes the proof.
\end{IEEEproof}

The next lemma will assist us to upper and lower lower bound the sum-rate in $\CoSTPILRed$. 
\begin{lemma}\label{Lem: Conv-Xk,n}
For $\lambda>0$, $\mu > 0$ and $\K$ growing to infinity at a rate $\K = \LO{N}$, we have
$$ 
\lim_{N \ra \infty} \frac{\tilde{R}\paren{\K,\lambda,\mu}}{\logp{\frac{1}{\lambda}\paren{\frac{1}{\beta_h}\logp{\alpha_h \K}}^\frac{1}{n_h}}}=1.
$$
\end{lemma}
\begin{IEEEproof}
Let $\Xrt{\lambda}{\mu}=\frac{\logp{\Xr{\lambda_h}{\mu}}}{\logp{\frac{1}{\lambda_h}\paren{\frac{1}{\beta_h}\logp{\alpha_h \K}}^\frac{1}{n_h}}}\I{\Xr{\lambda}{\mu}\geq 1}$. First, we show that $\Xrt{\lambda}{\mu}\xrightarrow{i.p.}1$ as $N$ grows large.  To this end, we will show that $g_{\pi(\K)}$ converges to zero in probability.  The CDF of the $\K $th smallest value for the collection of random variables $\left\{g_i\right\}_{i=1}^N$, which we denote as $F^{(\K )}_g(x)$, is given by
\begin{eqnarray}\label{Eq: CDF g}
F^{(\K )}_g(x)&=&\int_{0}^{F_g(x)}\frac{x^{\K -1}(1-x)^{N-\K }}{\text{B}\paren{\K ,N-\K +1}}dx,
\end{eqnarray}
where $\text{B}(a,b)=\int_{0}^1t^{a-1}(1-t)^{b-1}dt$ is the beta function \cite{orderstat}, and $F_g(x)$ is the CDF common to all $g_i$, $1\leq i\leq N$.  We define $\z{\K}$ as $\z{\K }= F_g\paren{g_{\pi(\K)}}$. Using \eqref{Eq: CDF g}, the CDF of $\z{\K }$, which we denote as $F_{\z{{\K} }}(x)$, is given as 
\begin{eqnarray}\label{Eq: CDF z}
F_{\z{{\K} }}(x)&=&\PR{\z{\K }\leq x}\nonumber\\
&\stackrel{\rm (a)}{=}&\PR{\g{\K }{N}\leq F^{-1}_g(x)}\nonumber\\
&=&\int_{0}^{x}\frac{x^{\K -1}(1-x)^{N-\K }}{\text{B}\paren{\K ,N-\K +1}}dx, 
\end{eqnarray}
where (a) follows from the fact that $F_g(x)$ is invertible and monotone increasing for $x>0$ and $g_{\pi(\K)} = \g{\K }{N}$.  Note that the random variable $X$ is said to be Beta distributed with parameters $v$ and $w$ if its CDF is given by $F_X\paren{x}=\int_{0}^x\frac{t^{v-1}\paren{1-t}^{w-1}}{\text{B}\paren{v,w}}dt$.  Thus, $\z{\K }$ is indeed a Beta distributed random variable with parameters $\K$ and $N-\K+1$.  Using the fact that $\z{\K }$ is Beta distributed, we can upper bound the tail probability $\PR{g_{\pi(\K)}>\epsilon}$ of $g_{\pi(\K)}$ for all $\epsilon>0$ as  
\begin{eqnarray}
\PR{g_{\pi(\K)}>\epsilon}&=&\PR{F_g\paren{g_{\pi\paren{\K}}}>F\paren{\epsilon}}\nonumber\\
&=&\PR{\z{\K}>F\paren{\epsilon}}\nonumber\\
&\stackrel{\rm (a)}{\leq}&\frac{\ES{\z{\K}}}{F\paren{\epsilon}}\nonumber\\
&\stackrel{\rm (b)}{=}&\frac{\K}{F\paren{\epsilon}\paren{N+1}},
\end{eqnarray}
where (a) follows from the Markov inequality, and (b) follows from the  formula $\ES{\z{\K}}=\frac{\K}{N+1}$ for the mean value of Beta distributed random variables \cite{Hastings}.  This implies $g_{\pi(\K)}\xrightarrow{i.p.}0$ as $N$ grows large.  We will use this convergence property of $g_{\pi(\K)}$ while we obtain a tight lower bound for $\Xr{\lambda}{\mu}$, hence for $\Xrt{\lambda}{\mu}$, below.  

We note that the collection of random variables $\brparen{h_{\pi\paren{i}}}_{i=1}^{\K}$ are i.i.d. with the common distribution $F_{h}\paren{x}$ because $\mathbi{h}$ and $\mathbi{g}$ are independent, and our selection criterion depends on $\mathbi{g}$.  Since $g_{\pi\paren{\K}}$ is the largest value in $\brparen{g_{\pi\paren{i}}}_{i=1}^{\K}$, $\Xr{\lambda}{\mu}$ can be lower bounded as $\max\limits_{1\leq i\leq \K }\frac{h_{\pi(i)}}{\lambda+\mu g_{\pi(\K)}}\leq \Xr{\lambda}{\mu}$. Therefore, we obtain the following  upper and lower bounds for $\Xr{\lambda}{\mu}$:
\begin{eqnarray}
\frac{\h{\K}}{\lambda+\mu g_{\pi\paren{\K}}}\leq \Xr{\lambda}{\mu}\leq \frac{\h{\K}}{\lambda},
\end{eqnarray} 
where $\h{\K}=\max_{1\leq i\leq \K} h_{\pi(i)}$. From continuous mapping theorem \cite{Billingsley}, we have $\frac{1}{1+\frac{\mu}{\lambda} g_{\pi\paren{\K}}}\xrightarrow{i.p.}1$ as $N$ grows large, which implies $\frac{\frac{1}{1+\frac{\mu}{\lambda} g_{\pi\paren{\K}}}\frac{\h{\K}}{\lambda}}{\frac{1}{\lambda}\paren{\frac{1}{\beta_h}\logp{\alpha_h \K}}^\frac{1}{n_h}}\xrightarrow{i.p.}1$ as $N$ grows large by using Lemma \ref{Lem: Conv-Xt}.  Thus, $\frac{\Xr{\lambda}{\mu}}{\frac{1}{\lambda}\paren{\frac{1}{\beta_h}\logp{\alpha_h \K}}^\frac{1}{n_h}}$ converges to $1$ in probability, which, in turn, implies the convergence of $\Xrt{\lambda}{\mu}$ to $1$ in probability.  The proof of uniform integrability is similar to that in Lemma \ref{Lem: Conv-Xt}.  
\end{IEEEproof}

In the next lemma, we establish an important convergence property for the Lagrange multiplier $\lambda_N$ as $N$ grows large.  This result will be used to obtain lower and upper bounds for the sum-rate in $\CoSTPILRed$ and to show the logarithmic effect of $P_{\rm ave}$ on the secondary network throughput under $\CoSTPILRed$.
\begin{lemma}\label{Lem: Lambda-Convergence}
Let $\lambda_N$ be the Lagrange multiplier corresponding to the average total transmit power constraint in $\CoSTPILRed$. Then, $\lim_{N\ra\infty}\lambda_N = \frac{1}{P_{\rm ave}}$. 
\end{lemma}
\begin{IEEEproof}
First, we show that $\liminf_{N\ra\infty}\lambda_N>0$ by contradiction.  Assume that $\liminf_{N\ra\infty}\lambda_N=0$.  This means that, for any given $\epsilon>0$, we can find a subsequence $\brparen{N_j}_{j=1}^\infty$ such that $\lambda_{N_j}\leq \epsilon$ for $N_j$ large enough.  Recall $\OPC{i^\star_{\K}}{\K}=\paren{\frac{1}{\lambda_{N}+\mu_{N} g_{i^\star_{\K}}}-\frac{1}{h_{i^\star_{\K}}}}^+$, where $i^\star_{\K}=\pi\paren{\arg\max\limits_{1\leq i\leq \K }{\frac{h_{\pi(i)}}{\lambda_N+\mu_Ng_{\pi(i)}}}}$.  The average total transmit power of the secondary network for $N_j$ large enough can be lower bounded as
\begin{eqnarray}\label{Eq: Power-LB}
\ES{\OPC{i^\star_{K_{N_j}}}{K_{N_j}}}&\stackrel{(a)}{\geq}&\ES{\paren{\frac{\frac{\h{K_{N_j} }}{\lambda_{N_j}+\mu_{N_j} g_{\pi(K_{N_j})}}-1}{\h{K_{N_j}}}}^+}\nonumber\\
&=&\ES{\paren{\frac{1}{\lambda_{N_j}+\mu_{N_j} g_{\pi(K_{N_j})}}-\frac{1}{\h{K_{N_j}}}}^+}\nonumber\\
&\ge&\ES{\paren{\frac{1}{\epsilon+\mu_{N_j}g_{\pi(K_{N_j})}}-\frac{1}{\h{K_{N_j}}}}^+},
\end{eqnarray}
where $\h{K_{N_j}}=\max_{1\leq i\leq K_{N_j} }h_{\pi(i)}$ and $(a)$ follows from the inequality that $g_{\pi(i)}\leq g_{\pi(K_{N_j})}$ for all $i \in \brparen{1, \ldots, K_{N_j}}$.  Note that the Lagrange multiplier for the average interference power can be upper bounded as $\mu_N \leq \frac{1}{Q_{\rm ave}}$ for all $N$, and also $\frac{1}{\h{K_{N_j}}}\xrightarrow{i.p}0$ as $N_j$ grows large.  Therefore, $\paren{\frac{1}{\epsilon+\mu_{N_j} g_{\pi(K_{N_j})}}-\frac{1}{\h{K_{N_j}}}}^+\xrightarrow{i.p}\frac{1}{\epsilon}$ as $N_j$ goes to infinity. Applying Fatou's Lemma to, we have
\begin{eqnarray}
\liminf_{N_j\ra\infty}\ES{\paren{\frac{1}{\lambda_{N_j}+\mu_{N_j} g_{i^\star_{K_{N_j}}}}-\frac{1}{h_{i^\star_{K_{N_j}}}}}^+}&\geq&\liminf_{N_j\ra\infty}\ES{\paren{\frac{1}{\epsilon+\mu_{N_j} g_{\pi(K_{N_j})}}-\frac{1}{\h{K_{N_j}}}}^+}\nonumber\\
&\geq&\frac{1}{\epsilon}.\nonumber
\end{eqnarray}
This implies that the average total transmit power constraint will be violated when $N_j$ is large enough if we choose $\epsilon$ sufficiently small.  Thus, we conclude that $\liminf_{N\ra\infty}\lambda_N>0$. 

Now, we will conclude the proof by using the fact that $\lambda_N$ cannot be arbitrarily close to zero.  The average total transmit power constraint can be written as $P_{\rm ave}=\ES{\paren{\frac{1}{\lambda_N+\mu_N g_{i^\star_{\K}}}-\frac{1}{h_{i^\star_{\K}}}}^+}$.
This implies $\lambda_N\leq \frac{1}{P_{\rm ave}}$ for all $N$, and hence $\limsup_{N\ra\infty}\lambda_N\leq \frac{1}{P_{\rm ave}}$. Following similar steps that we used to drive \eqref{Eq: Power-LB}, $\lambda_N$ can be lower bounded as $\lambda_N\geq \frac{1}{P_{\rm ave}}\ES{\paren{\frac{1}{1+\frac{\mu_N}{\lambda_N} g_{\pi\paren{\K}}}-\frac{\lambda_N}{\h{\K}}}^+}$.
Since $\liminf_{N\ra\infty}\lambda_N>0$ and $\mu_N\leq \frac{1}{Q_{\rm ave}}$, we have $\frac{1}{1+\frac{\mu_N}{\lambda_N} g_{\pi\paren{\K}}}\xrightarrow{i.p.}1$ as $N$ goes to infinity ({\em i.e.,} see the proof of Lemma \ref{Lem: Conv-Xk,n} for the convergence of $g_{\pi\paren{\K}}$ to $0$ in probability).  We also have $\frac{\lambda_N}{\h{\K}}\xrightarrow{i.p.}0$ as $N$ goes to infinity because $\lambda_N\leq \frac{1}{P_{\rm ave}}$. Thus, $\paren{\frac{1}{1+\frac{\mu_N}{\lambda_N} g_{\pi\paren{\K}}}-\frac{\lambda_N}{\h{\K}}}^+$ converges to $1$ in probability.  Applying Fatou's Lemma, we have
\begin{eqnarray}
\liminf_{N\ra\infty}\lambda_N&\geq& \frac{1}{P_{\rm ave}}\liminf_{N\ra\infty}\ES{\paren{\frac{1}{1+\frac{\mu_N}{\lambda_N} g_{\pi\paren{\K}}}-\frac{1}{\h{\K}}}^+}\nonumber\\
&\geq& \frac{1}{P_{\rm ave}},
\end{eqnarray}
which completes the proof.
\end{IEEEproof}


\subsection{Proof of Throughput Scaling in $\CoSTPILRed$}
We first note that $\tilde{R}\paren{\K, \lambda, \mu}$ is a decreasing function of $\lambda$ and $\mu$.  Thus, for any given $\epsilon>0$, we can find a constant $N_0$ large enough such that $\RateTPILRed$ can be upper and lower bounded as
\begin{eqnarray}\label{Eq: Xr,n-up-low-bound}
 \tilde{R}\paren{\K,\frac{1+\epsilon}{P_{\rm ave}},\frac{1}{Q_{\rm ave}}}\leq \RateTPILRed \leq \tilde{R}\paren{\K,\frac{1-\epsilon}{P_{\rm ave}},0}
\end{eqnarray}
for all $N \geq N_0$ since $\lambda_N$ converges to $\frac{1}{P_{\rm ave}}$ by Lemma \ref{Lem: Lambda-Convergence} and $\mu_N \leq \frac{1}{Q_{\rm ave}}$.  
Using Lemma \ref{Lem: Conv-Xk,n}, for any given $\epsilon>0$ and $N$ large enough, $\RateTPILRed$ can be further upper and lower bounded as 
\begin{eqnarray}\label{Eq: FinalTPILRed}
\lefteqn{\paren{1-\epsilon}\left[\logp{\frac{P_{\rm ave}}{1+\epsilon}}+\frac{1}{n_h}\logp{\frac{1}{\beta_h}\logp{\alpha_h \K}}\right]} \hspace{16cm} \nonumber \\
\lefteqn{\leq \RateTPILRed \leq  \paren{1+\epsilon}\left[\logp{\frac{P_{\rm ave}}{1-\epsilon}}+\frac{1}{n_h}\logp{\frac{1}{\beta_h}\logp{\alpha_h \K}}\right],} \hspace{10.7cm} 
\end{eqnarray}
which implies 
\begin{eqnarray}
\lim_{N\ra\infty}\frac{\RateTPILRed}{\logp{\logp{ \K}}}=\frac{1}{n_h}.\nonumber
\end{eqnarray}
\subsection{Proof of Throughput Scaling in $\CoSTPILFull$}
Consider a secondary network under $\CoSTPILFull$ with a total average transmit power constraint $P_{\rm ave}$, total average interference power constraint $Q_{\rm ave}$ and $N$ SUs.  By removing the average interference power constraint, we obtain a primary MAC network with a total average transmit power constraint $P_{\rm ave}$.  Hence, the sum-rate in $\CoSTPILFull$ can be upper bounded by the sum-rate $ \RateTPM$ of the primary MAC network with the same total average transmit power constraint, {\em i.e.,} $\RateTPILFull \leq \RateTPM$.  In the next lemma, we establish the asymptotic scaling behavior of $\RateTPM$, which will also serve as an upper bound for $\RateTPILFull$.  
\begin{lemma}\label{Lem: TPM}
$\RateTPM$ scales according to 
$
\lim_{N\ra\infty}\frac{\RateTPM}{\log\logp{N}}=\frac{1}{n_h}
$.  Furthermore, for any given $\epsilon>0$, there exists a constant $N_0$ large enough such that \begin{eqnarray}
\paren{1-\epsilon}\left[\logp{\frac{P_{\rm ave}}{1+\epsilon}}+\frac{1}{n_h}\logp{\frac{1}{\beta_h}\logp{\alpha_h N}}\right] \leq \RateTPM \leq  \paren{1+\epsilon}\left[\logp{\frac{P_{\rm ave}}{1-\epsilon}}+\frac{1}{n_h}\logp{\frac{1}{\beta_h}\logp{\alpha_h N}}\right]\nonumber
\end{eqnarray}
for all $N \geq N_0$. 
\end{lemma}
\begin{IEEEproof}
Since the proof is similar to the analysis given above, we skip it to avoid repetitions. 
\end{IEEEproof}

The proof of throughput scaling in $\CoSTPILFull$ is completed if we obtain a lower bound for $\RateTPILFull$ that also scales according to $\frac{1}{n_h} \log\logp{N}$ as $N$ grows large.  To this end, we observe that $\RateTPILRed$ serves as a lower bound for $\RateTPILFull$ since more information is available at the SBS to perform power control and user scheduling under $\CoSTPILFull$.  Thus, for any given $\epsilon>0$, $0<\delta<1$ and $N$ large enough, we have
\begin{eqnarray}\label{Eq: FinalTPILFull}
\lefteqn{\paren{1-\epsilon}\left[\logp{\frac{P_{\rm ave}}{1+\epsilon}}+\frac{1}{n_h}\logp{\frac{1}{\beta_h}\logp{\alpha_h N^\delta}}\right]} \hspace{16cm} \nonumber \\
\lefteqn{\leq \RateTPILFull \leq  \paren{1+\epsilon}\left[\logp{\frac{P_{\rm ave}}{1-\epsilon}}+\frac{1}{n_h}\logp{\frac{1}{\beta_h}\logp{\alpha_h N}}\right],} \hspace{10.2cm} 
\end{eqnarray} 
which implies
\begin{eqnarray}
\lim_{N\ra\infty}\frac{\RateTPILFull}{\logp{\logp{ N}}}=\frac{1}{n_h}. 
\end{eqnarray}
\section{Scaling Behavior of Interference in Total-Power-And-Interference-Limited Scenario}\label{App: Interference-Conv}
Since $\mathcal{I}_{\K}$ is a positive random variable, it is enough to show that $\lim_{N\ra\infty}\ES{\mathcal{I}_{\K}}=0$ for $\K$ increasing to infinity at a rate $\K =\LO{N}$ as $N$ goes to infinity.  We can upper bound $\mathcal{I}_{\K}$ as 
\begin{eqnarray}
\mathcal{I}_{\K}&=&g_{i^\star_{\K}}\OPC{i^\star_{\K}}{\K}\nonumber\\
&=&g_{i^\star_{\K}}\paren{\frac{1}{\lambda_N + \mu_N g_{i^\star_{\K}}} -  \frac{1}{h_{i^\star_{\K}}}}^{+} \nonumber\\
&\leq&\frac{\g{\K}{N}}{\lambda_N}.\nonumber
\end{eqnarray}
Hence, using the fact that $\liminf_{N \ra\infty}{\lambda_N} > 0$, it is enough to show that $\lim_{N\ra\infty}\ES{\g{\K}{N}}=0$.  To this end, for any given $\epsilon>0$, we can upper bound $\ES{\g{\K}{N}}$ as 
\begin{eqnarray}\label{Eq: exp-g-upb}
\ES{\g{\K}{N}}&=&\ES{\g{\K}{N}\I{\g{K}{N}< \epsilon}}+\ES{\g{\K}{N}\I{\epsilon \leq \g{K}{N}< N}}+\sum_{i=1}^\infty\ES{\g{\K}{N}\I{N^i\leq \g{K}{N} < N^{i+1}}}\nonumber\\
&\leq & \epsilon + N\PR{\g{\K}{N}\geq \epsilon}+\sum_{i=1}^\infty N^{i+1}\PR{\g{\K}{N}\geq N^i}. 
\end{eqnarray}

Below, we will show that $\PR{\g{\K}{N}\geq \epsilon}$ and $\PR{\g{\K}{N}\geq N^i}$ can be asymptotically bounded as $\PR{\g{\K}{N}\geq \epsilon} \leq \e{-\BT{N}}$ and $\PR{\g{\K}{N}\geq N^i} \leq \e{-\BT{N^{in+1}}}$.  Assuming for a while that these asymptotic bounds hold, $\ES{\g{\K}{N}}$ can be upper bounded as
\begin{eqnarray}
\ES{\g{\K}{N}}&\leq & \epsilon + N\e{-\BT{N}}+\sum_{i=1}^\infty N^{i+1}\e{-\BT{N^{in+1}}}\nonumber\\
& \leq & \epsilon + N\e{-\BT{N}}+N\e{-N}\sum_{i=1}^\infty N^{i}\e{-\BT{N^{in}}}, \nonumber
\end{eqnarray}
which implies $\limsup_{N\ra\infty}\ES{\g{\K}{N}}=0$ and completes the proof.

Now, we derive the aforementioned asymptotic expansions for $\PR{\g{\K}{N}\geq \epsilon}$ and $\PR{\g{\K}{N}\geq N^i}$.  Note that $\PR{\g{\K}{N}\leq x}=\sum_{i=\K}^N {N \choose i}F_g^i\paren{x}\paren{1-F_g\paren{x}}^{N-i}$ \cite{orderstat}. Therefore, for all $x>0$ and $N$ large enough, $\PR{\g{\K}{N} \geq x}$ can be upper bounded as
\begin{eqnarray}\label{Eq: Order-Statistics-Tail}
\PR{\g{\K}{N} \geq x} 
&=&\sum_{i=0}^{\K-1} {N \choose i}F_g^i\paren{x}\paren{1-F_g\paren{x}}^{N-i}\nonumber\\
&\stackrel{(a)}{\leq} & \K {N \choose \K}\paren{1-F_g\paren{x}}^{N-\K}\nonumber\\
&\stackrel{(b)}{\leq}& \K \sqrt{\frac{N}{\pi \K\paren{N-\K}}}2^{ NH_b\paren{\frac{\K}{N}}}\paren{1-F_g\paren{x}}^{N-\K}\nonumber\\
&\leq & N 2^{ NH_b\paren{\frac{\K}{N}}}\paren{1-F_g\paren{x}}^{N-\K},
\end{eqnarray}
where $\paren{a}$ follows from the fact that $\K<\paren{N+1}F_g\paren{x}-1$ for $x>0$ and $N$ large enough\footnote{Let $b\paren{k; n, p}$ be the Binomial function with parameters $n \in \N$ and $p \in \paren{0, 1}$, which is defined as $b\paren{k;n,p}={n\choose k}p^k \paren{1-p}^{n-k}$.  Then, $b\paren{k;n,p}$ first increases, and then decreases as a function of $k$, reaching its maximum at $k=m^\star$, where $m^\star$ is an integer such that $\paren{n+1}p-1<m^\star\leq \paren{n+1}p$.  If $\paren{n+1}p$ is an integer, then $m^\star=\paren{n+1}p$, and $b\paren{m^\star-1;n,p}=b\paren{m^\star;n,p}$ \cite{Feller-Probability}.}, $\paren{b}$ follows from ${N\choose \K}\leq2^{NH_b\paren{\frac{\K}{N}}} \sqrt{\frac{N}{\pi \K\paren{N-\K}}}$ for $\K\notin\brparen{0,N}$, where $H_b(\cdot)$ is the binary entropy function \cite{Cover}.
 Using \eqref{Eq: Order-Statistics-Tail}, $\PR{\g{\K}{N}\geq \epsilon}$ can be upper bounded as 
\begin{eqnarray}\label{Eq: exp-g-upb-1}
\PR{\g{\K}{N}\geq \epsilon}&\leq & N 2^{ NH_b\paren{\frac{\K}{N}}}\paren{1-F_g\paren{\epsilon}}^{N-\K}\nonumber\\
&= &\e{N\paren{\paren{1-\frac{\K}{N}}\logp{1-F_g\paren{\epsilon}}+H_b\paren{\frac{\K}{N}}\logp{2}+\frac{\logp{N}}{N}}}\nonumber\\
&=&\e{\BT{-N}}.
\end{eqnarray}

Similarly, for any given $\epsilon>0$ and $N$ large enough, $\PR{\g{\K}{N}\geq N^i}$ can be upper bounded as 
\begin{eqnarray}\label{Eq: exp-g-upb-2}
\PR{\g{\K}{N}\geq N^i}&\leq&  N 2^{ NH_b\paren{\frac{\K}{N}}}\paren{1-F_g\paren{N^i}}^{N-\K}\nonumber\\
&\leq& N 2^{ NH_b\paren{\frac{\K}{N}}}\paren{\paren{1+\epsilon}\alpha_h N^{il_h}\e{-\beta_h N^{in_h}+H\paren{N^i}}}^{N-\K}\nonumber\\
&=& \e{\paren{\logp{N}+ NH_b\paren{\frac{\K}{N}}\logp{2}+\paren{N-\K}\paren{\log\paren{1+\epsilon}+\logp{\alpha_h N^{il_h}}-\beta_h N^{in_h}+H\paren{N^i}}}}\nonumber\\
&=& \e{N^{in_h+1}\paren{\paren{1-\frac{\K}{N}}\paren{-\beta_h+\frac{\log\paren{1+\epsilon}+\logp{\alpha_h N^{il_h}} +H\paren{N^i}}{N^{in_h}}}+\frac{\logp{N}+ NH_b\paren{\frac{\K}{N}}\logp{2}}{N^{in_h+1}}}}\nonumber\\
&=&\e{\BT{-N^{in+1}}}.
\end{eqnarray}
\section{Throughput Scaling in Interference-Limited Scenario}\label{App: IL}
In this appendix, we first establish the secondary network throughput scaling behavior for $\CoSILFull$.  Then, we will use this result to obtain an upper bound on the secondary network throughput in $\CoSILRed$.  Without loss of generality, we assume that $W$, \emph{i.e.,} interference plus noise power, is equal to 1, and establish the throughput scaling behavior of IL networks.

To this end, we need to study the asymptotic behavior of $\X{0}{\mu_N} = \max_{1\leq i\leq N}\frac{h_i}{\mu_Ng_i}$ when $N$ grows large.  Similar to our proof for TPIL networks given in Appendix \ref{App: TPIL}, we start our analysis by deriving the scaling behavior of  $\tilde{R}\paren{N, 0, \mu} = \ES{\logp{\X{0}{\mu}}\I{\X{0}{\mu}\geq1}}$, where $\mu$ is a fixed positive constant.  In the next lemma, we characterize the asymptotic tail behavior of the random variable $X_i\paren{0, \mu} = \frac{h_i}{\mu g_i}$, where $X_i\paren{\lambda, \mu}$ is defined as $X_i\paren{\lambda, \mu} = \frac{h_i}{\lambda+\mu g_i}$.  Then, we will use this lemma to establish the concentration behavior of $\X{0}{\mu}$, and thereby to obtain the scaling behavior of $\tilde{R}\paren{N, 0, \mu}$. Throughout this appendix, we assume that $h_i$'s and $g_i$'s are distributed according to $F_h\paren{x}$ and $F_g\paren{x}$, respectively, where $F_h\paren{x}$ and $F_g\paren{x}$ belong to the class $\mathcal{C}$ distributions. 
\begin{lemma}\label{Lem: IL-tail}
Let $F_{X_i\paren{0, \mu}}(x)$ be the CDF of $X_i\paren{0, \mu} = \frac{h_i}{\mu g_i}$. Then,
$
\lim_{x\ra\infty}\frac{1-F_{X_i\paren{0, \mu}}(x)}{\xi x^{-\gamma_g}}=1
$, where $\xi = \frac{\eta_g\ES{h_i^{\gamma_g}}}{\mu^{\gamma_g}}$, and $\eta_g$ and $\gamma_g$ are positive constants derived from the behavior of the distribution function of $g_i$ near the origin. 
\end{lemma}
\begin{IEEEproof}
To prove this lemma, we use some results from \cite{Andrey} characterizing the asymptotic tail behavior for the product of two independent random variables.  Let $U$ and $V$ be two independent random variables with distribution functions $F_U\paren{x}$ and $F_V\paren{x}$, respectively.  The asymptotic tail behavior of the distribution function $F_{UV}(x)$ of $UV$ is given by $\lim_{x \ra \infty} \frac{1-F_{UV}(x)}{C \ES{U^\theta} x^{-\theta}} = 1$ if the following conditions hold ({\em i.e.,} Theorem 4 in \cite{Andrey}): \emph{i-)} $\lim_{x \ra \infty} \frac{1-F_V(x)}{C x^{-\theta}} = 1$ for some positive constants $C>0$ and $\theta>0$, \emph{ii-)} $1-F_U\paren{\phi\paren{x}}=\LO{x^{-\theta}}$ as $x$ tends to infinity for some $\phi\paren{x}$ such that $\phi\paren{x} = \LO{x}$ as $x$ grows large, and \emph{iii-)} $\ES{U^\theta}<\infty$. To derive the asymptotic behavior of $\frac{h_i}{\mu g_i}$, it is enough to show that $\frac{1}{\mu g_i}$ and $h_i$ satisfy these conditions.  Since $F_g(x)$ varies regularly near the origin, the tail of the distribution function of $\frac{1}{\mu g_i}$ varies regularly, {\em i.e.}, $\lim_{x \ra \infty}\frac{\PR{\frac{1}{\mu g_i} > x}}{\eta_g \paren{\mu x}^{-\gamma_g}}=1$, implying $C = \frac{\eta_g}{\mu^{\gamma_g}}$ and $\theta = \gamma_g$ above.  Choosing $\phi(x)$ as $\phi(x)=x^\delta$ for some $\delta \in \paren{0, 1}$, we have $1-F_h\paren{x^\delta}=\BO{\alpha_h x^{\delta l_h}\e{\paren{-\beta_h x^{\delta n_h}+H\paren{x^\delta}}}}=\LO{x^{-\gamma_g}}$.  Since the tail of $1-F_h(x)$ decays exponentially to zero, it also follows that $\ES{h_i^{\gamma_g}}<\infty$, which completes the proof. 
\end{IEEEproof}

Now, we provide a key lemma that will enable us to upper and lower bound the secondary network scaling behavior in $\CoSILFull$.  
\begin{lemma}\label{Lem: Yt-scaling-IL}
For $\mu>0$, we have $\lim_{N\ra\infty}\frac{\tilde{R}\paren{N, 0, \mu}}{\frac{1}{\gamma_g}\logp{\xi N}}=1$. 
\end{lemma}
\begin{IEEEproof}
Let $\Xt{0}{\mu}=\frac{\logp{\X{0}{\mu}}}{\frac{1}{\gamma_g}\logp{\xi N}}\I{\X{0}{\mu}\geq1}$ where $\X{0}{\mu}=\max_{1\leq i\leq N}X_i\paren{0,\mu}$.  It is sufficient to show that $\lim_{N\ra\infty}\ES{\Xt{0}{\mu}}=1$. We start our analysis by proving that $\Xt{0}{\mu}\xrightarrow{i.p.}1$ as $N$ tends to infinity.  Using Lemma \ref{Lem: IL-tail}, the tail behavior of $F_{X_i\paren{0,\mu}}\paren{x}$ is characterized by $G\paren{x}=\frac{x^{\gamma_g}}{\xi}$, where $\xi = \frac{\eta_g\ES{h_i^{\gamma_g}}}{\mu^{\gamma_g}}$.  Hence, $\inv{G}{x}=\paren{\xi x}^\frac{1}{\gamma_g}$. Using Lemma \ref{Lem: Concentration}, we have
\begin{eqnarray}
\PR{\inv{G}{\xi^{-\epsilon}N^{1-\epsilon}}<\X{0}{\mu}\leq \inv{G}{\xi^{\epsilon}N^{1+\epsilon}}}&=&\PR{\paren{\xi N}^{\frac{1}{\gamma_g}\paren{1-\epsilon }}< \X{0}{\mu}\leq \paren{\xi N}^{\frac{1}{\gamma_g}\paren{1+\epsilon }}}\nonumber\\
&=&1-\BT{\frac{1}{N^{\epsilon} }},
\end{eqnarray}
which implies $\Xt{0}{\mu}\xrightarrow{i.p.}1$ as $N$ grows large. Since convergence in probability does not always imply convergence in mean, we need to show that $\left\{\Xt{0}{\mu}\right\}_{N=1}^{\infty}$ is uniformly integrable to complete the proof.  To this end, we will show that for all $\epsilon_1>0$ and $\epsilon_2>0$, there exist large enough positive constants $C_0$ and $N_0$ such that
\begin{eqnarray}\label{Eq: IL-integrability-upb}
\sup_{N\geq N_0}\ES{\Xt{0}{\mu}\I{\Xt{0}{\mu}> C}}&\leq&\paren{1+\epsilon_1}\paren{1+\epsilon_2}\frac{\xi N_0}{\paren{\xi N_0}^{C}}\sum_{i=0}^{\infty}\frac{\paren{i+2}}{\paren{\xi N_0}^{iC}},
\end{eqnarray}
for $C\geq C_0$.  Since the proof of this fact is similar to the proof of \eqref{Eq: sup-upperbound}, we just give the main steps below. 

For $C>0$, $\ES{\Xt{0}{\mu}\I{\Xt{0}{\mu}> C}}$ can be upper bounded as
\begin{eqnarray}
\ES{\Xt{0}{\mu}\I{\Xt{0}{\mu}> C}} 
&\leq& \sum_{i=1}^{\infty}\paren{i+1}C\PR{iC< \Xt{0}{\mu}\leq\paren{i+1}C}\nonumber\\
&\leq&\sum_{i=1}^{\infty}\paren{i+1}C\PR{\paren{\xi N}^{\frac{iC}{\gamma_g}}< \X{0}{\mu}}.
\end{eqnarray}
For $\epsilon_1>0$, $\epsilon_2>0$ and $N$ large enough, $\PR{\paren{\xi N}^{\frac{iC}{\gamma_g}}<\X{0}{\mu}}$ can be upper bounded as
\begin{eqnarray}
\PR{\paren{\xi N}^{\frac{iC}{\gamma_g}}<\X{0}{\mu}} 
&\stackrel{(a)}{\leq}&1-\paren{1-\frac{\xi \paren{1+\epsilon_1}}{\paren{\xi N}^{iC}}}^N\nonumber\\
&=&1-\e{N\logp{1-\frac{\xi \paren{1+\epsilon_1}}{\paren{\xi N}^{iC}}}}\nonumber\\
&\stackrel{(b)}{\leq}& 1-\e{-\paren{1+\epsilon_1}\paren{1+\epsilon_2}\frac{\xi N}{\paren{\xi N}^{iC}}}.\nonumber\\
&\stackrel{(c)}{\leq} &\paren{1+\epsilon_1}\paren{1+\epsilon_2}\frac{\xi N}{\paren{\xi N}^{iC}},
\end{eqnarray}
where $(a)$ follows from the fact that for all $\epsilon_1>0$ and $x$ large enough, $F_{X\paren{0,\mu}}(x)$ can be lower bounded as $F_{X\paren{0,\mu}}(x) \geq 1-\frac{\xi \paren{1+\epsilon_1}}{x^{\gamma_g}}$ (recall $\lim_{x\ra\infty}\frac{1-F_{X\paren{0,\mu}}(x)}{\xi x^{-\gamma_g}}=1$), $(b)$ follows form the fact that for all $\epsilon_2>0$ and $x$ small enough, $\logp{1- x}$ can be lower bounded as $\logp{1-x} \geq -\paren{1+\epsilon_2}x$ since $\lim_{x\downarrow0}\frac{\logp{1-x}}{x}=-1$, and $(c)$ follows from the fact $\e{-x}$ can be lower bounded as $\e{-x}\geq 1-x$ for $x>0$. 

Note that $\sum_{i=1}^{\infty}\paren{i+1}C\paren{1+\epsilon_1}\paren{1+\epsilon_2}\frac{\xi N}{\paren{\xi N}^{iC}}$ is a decreasing function of $N$ for $C>1$.  Thus, for given $\epsilon_1>0$ and $\epsilon_2>0$, we can find large enough constants $N_0$ and $C_0$ such that 
\begin{eqnarray}
\sup_{N\geq N_0}\ES{\Xt{0}{\mu}\I{\Xt{0}{\mu}> C}} 
&\leq& C\paren{1+\epsilon_1}\paren{1+\epsilon_2}\frac{\xi N_0}{\paren{\xi N_0}^{C}}\sum_{i=0}^{\infty}\frac{\paren{i+2}}{\paren{\xi N_0}^{iC}},\nonumber
\end{eqnarray}
 for $C\geq C_0$.
\end{IEEEproof}

The next lemma characterizes the asymptotic behavior of the Lagrange multiplier $\mu_N$ as $N$ becomes large.  Later, we will use this result to provide upper and lower bounds for $\tilde{R}\paren{N,0,\mu}$. This result will also be helpful to conclude the logarithmic effect of $Q_{\rm ave}$ on the secondary network throughput under $\CoSILFull$.
 \begin{lemma}\label{Lem: mu convergence}
Let $\mu_N$ be the Lagrange multiplier corresponding to the average interference power constraint in $\CoSILFull$.  Then, $\lim_{N \ra \infty} \mu_{N}=\frac{1}{Q_{\rm ave}}$.
 \end{lemma}
 \begin{IEEEproof}
First, we show that $\liminf_{N\ra\infty}\mu_N>0$ by contradiction.  Assume $\liminf_{N\ra\infty}\mu_N=0$. This means that, for any given $\epsilon>0$, we can find a subsequence $\brparen{N_j}_{j=1}^\infty$ such that $\mu_{N_j}\leq \epsilon$ for $N_j$ large enough.  The average interference power for $N_j$ large enough can be lower bounded as $\ES{\paren{\frac{1}{\mu_{N_j}}-\frac{1}{X^\star_{N_j}\paren{0,1}}}^+} \geq \ES{\paren{\frac{1}{\epsilon}-\frac{1}{X^\star_{N_j}\paren{0,1}}}^+}$.
Note that $\paren{\frac{1}{\epsilon}-\frac{1}{X^\star_{N_j}\paren{0,1}}}^+\xrightarrow{i.p.}\frac{1}{\epsilon}$. Applying Fatou's lemma , we have
 \begin{eqnarray}
\liminf_{N_j\ra\infty}\ES{\paren{\frac{1}{\mu_{N_j}}-\frac{1}{X^\star_{N_j}\paren{0,1}}}^+}&\geq& \liminf_{N_j\ra\infty}\ES{\paren{\frac{1}{\epsilon}-\frac{1}{X^\star_{N_j}\paren{0,1}}}^+}\nonumber\\
&\geq& \frac{1}{\epsilon},\nonumber
\end{eqnarray}
which implies that the average interference power constraint will be violated when $N_j$ is large enough if we choose $\epsilon$ sufficiently small.  Thus, we conclude that $\liminf_{N\ra\infty}\mu_N>0$.  Now, we will complete the proof by using the fact that $\mu_N$ cannot be arbitrarily close to zero as $N$ grows large.  The average interference power constraint can be expressed as $\mu_N=\frac{1}{Q_{\rm ave}}\ES{\paren{1-\frac{1}{\X{0}{\mu_N}}}^+}$.  Since $\liminf_{N\ra\infty}\mu_N>0$, the desired result follows from the dominated convergence theorem.\footnote{Note that Lebesgue dominated convergence theorem remains valid if almost sure convergence is replaced with convergence in probability in its hypothesis \cite{Bogachev}.}
 \end{IEEEproof}

\subsection{Proof of Throughput Scaling in $\CoSILFull$}
Now, we establish the secondary network throughput scaling behavior under the communication scenario $\CoSILFull$.  First, we note that $\tilde{R}\paren{N, 0, \mu}$ is a decreasing function of $\mu$.  Therefore, for any given $\epsilon>0$, we can find a constant $N_0$ large enough such that $\RateILFull$ can be upper and lower bounded as 
\begin{eqnarray}\label{Eq: Rate-IL-ulb}
\tilde{R}\paren{N,0,\frac{1+\epsilon}{Q_{\rm ave}}}\leq \RateILFull \leq \tilde{R}\paren{N,0,\frac{1-\epsilon}{Q_{\rm ave}}}
\end{eqnarray}
for all $N\geq N_0$ since $\mu_N$ converges to $\frac{1}{Q_{\rm ave}}$ as $N$ becomes large. Using Lemma \ref{Lem: Yt-scaling-IL}, for any $\epsilon>0$ and $N$ large enough, $\RateILFull$ can be further upper and lower bounded as
 \begin{eqnarray}\label{Eq: FinalILFull}
\paren{1-\epsilon}\frac{1}{\gamma_g}\logp{\paren{\frac{Q_{\rm ave}}{1+\epsilon}}^{\gamma_g} \eta_g \ES{h^{\gamma_g}} N} \leq \RateILFull \leq  \paren{1+\epsilon}\frac{1}{\gamma_g}\logp{\paren{\frac{Q_{\rm ave}}{1-\epsilon}}^{\gamma_g} \eta_g \ES{h^{\gamma_g}} N},
\end{eqnarray} 
where $h$ is a generic random variable with CDF $F_h(x)$.  Thus, we have
\begin{eqnarray}
\lim_{N\ra\infty}\frac{\RateILFull}{\logp{N}}=\frac{1}{\gamma_g}.
\end{eqnarray}
 \subsection{Proof of Throughput Scaling in $\CoSILRed$}
The secondary network sum-rate in $\CoSILFull$ serves as an upper bound for the secondary network sum-rate in $\CoSILRed$, \emph{i.e.,} $\RateILRed\leq \RateILFull$, since more information is available at the SBS to perform power control and user scheduling under $\CoSILFull$.  Thus, we have $\limsup_{N\ra\infty} \frac{\RateILRed}{\logp{N}}\leq \frac{1}{\gamma_g}$. 

To prove the other direction, consider a sub-optimum power allocation policy $\SPAP{IL}{\K}\paren{\vec{g}}$, which only depends on interference channel power gains, for a secondary network under $\CoSILRed$ in which SU-$i$ transmits with power $\SPC{IL}{\K}\paren{\vec{g}} = \frac{Q_{\rm ave}}{g_i}\I{g_i = \min_{1\leq i\leq \K}g_{\pi(i)}}$.  Note that $\min_{1\leq i\leq \K}g_{\pi(i)} =\min_{1\leq i\leq N}g_i = g_{\min}(N)$.  The average interference power at the PBS using $\SPAP{IL}{\K}$ is given by $\ES{\sum_{i=1}^N\SPC{IL}{\K}g_i}=Q_{\rm ave}$.
 Thus, $\SPAP{IL}{\K}$ satisfies the average interference power constraint, and therefore it is a feasible power allocation policy for a secondary network under $\CoSILRed$.  Let $\SRate{IL}{\K}$ be the secondary network sum-rate using $\SPAP{IL}{\K}$. $\hat{R}_{\rm IL}\paren{\K}$ can be lower bounded as  
 \begin{eqnarray}
\SRate{IL}{\K}
 &\geq&\logp{Q_{\rm ave}}+\ES{\logp{h}}+\ES{\logp{\frac{1}{g_{\min}(N)}}},\nonumber
 \end{eqnarray}
where $h$ is a generic random variable with CDF $F_h(x)$.  Using arguments similar to the ones used to prove Lemma \ref{Lem: Yt-scaling-IL}, it is easy to show that $\lim_{N\ra\infty}\frac{\ES{\logp{\frac{1}{g_{\min}(N)}}}}{\logp{N}}=\frac{1}{\gamma_g}$ since $\frac{1}{g_{\min}(N)}=\max_{1\leq i\leq N}\frac{1}{g_i}$ and the tail behavior of $\frac{1}{g_i}$ is characterized by $\lim_{x\ra\infty}\frac{\PR{\frac{1}{g_i}\geq x}}{\eta_g x^{-\gamma_g}}=1$. Also, we have  $\abs{\ES{\logp{h}}}<\infty$\footnote{Using Jensen inequality, we have $\ES{\logp{h}}\leq \logp{\ES{h}}=0$. Also, $\ES{\logp{h}}$ can be lower bounded as $\ES{\logp{h}}\geq \ES{\logp{h}\I{ h\leq 1}}$. Using integration by part and the fact that $F_g\paren{x}$ varies regularly around origin, it is easy to show that $\abs{\ES{\logp{h}\I{ h\leq 1}}}<\infty$.}. Thus, $\liminf_{N\ra\infty}\frac{\SRate{IL}{\K}}{\logp{N}}\geq\frac{1}{\gamma_g}$. Since $\SPAP{IL}{\K}$ is a sub-optimal power allocation policy for a secondary network under $\CoSILRed$, we have $\SRate{IL}{\K} \leq \RateILRed$ and therefore $\liminf_{N\ra\infty}\frac{\RateILRed}{\logp{N}}\geq\frac{1}{\gamma_g}$, which completes the proof.
\section{Throughput Scaling in Individual-Power-And-Interference-Limited Scenario}\label{App: IPIL}
In this appendix, we first establish the secondary network throughput scaling behavior for $\CoSIPILFull$. Then, we use this result to obtain an upper bound on the secondary network throughput scaling in $\CoSIPILRed$.  Below, we prove two key lemmas that will facilitate the proof of Theorem \ref{Theo: IPIL}. Without loss of generality, it is assumed that the interference plus noise power is equal to one.  The next lemma establishes the throughput scaling behavior for a primary MAC network. This result in turn leads to an upper bound on the sum-rate for $\CoSIPILFull$.
\begin{lemma}\label{Lem: Primary-MAC}
Consider a primary MAC network containing $N$ users with symmetric individual average transmit power constraints equal to $P_{\rm ave}$ and i.i.d. channel power gains $\brparen{h_i}_{i=1}^N$ whose common CDF belongs to the class $\mathcal{C}$ distributions.  Let $\RateIPM$ be the sum-rate of this network when transmit powers of users are optimally allocated.  Then, 
$
\lim_{N\ra\infty}\frac{\RateIPM}{\logp{N}}=1.
$
\end{lemma}  
\begin{IEEEproof}
$\RateIPM$ can be expressed as $ \RateIPM=\ES{\log\paren{\frac{\h{N}}{\lambda_N}}\I{\frac{\h{N}}{\lambda_N}\geq1}}$, where $\lambda_N$ is the Lagrange multiplier associated with individual power constraints.  As above, we define $\h{N}$ as $\h{N}=\max_{1\leq i\leq N}{h_i}$.  $\RateIPM$ can be upper and lower bounded as 
\begin{eqnarray}
\logp{\frac{1}{\lambda_N}}+\ES{\logp{\h{N}}} \leq \RateIPM \leq \log\paren{\frac{1}{\lambda_N}}+\ES{\log\paren{\h{N}}\I{\h{N}\geq1}}. \nonumber
\end{eqnarray}

Below, we will show that $\lim_{N \ra \infty} N\lambda_N = \frac{1}{P_{\rm ave}}$.  This will complete the proof since $\ES{\logp{\h{N}}}$ scales double logarithmically with $N$ ({\em i.e.,} see Lemma \ref{Lem: Conv-Xt} in Appendix \ref{App: TPIL}).  
The average transmit power of the $i$th user can be written as
\begin{eqnarray}\label{Eq: Pave}
\ES{\paren{\frac{1}{\lambda_N}-\frac{1}{h_i}}^+\I{h_i=\h{N}}} 
 &=&\frac{1}{N}\ES{\paren{\frac{1}{\lambda_N}-\frac{1}{\h{N}}}\I{\h{N}\geq\lambda_N}}\nonumber\\  
 &=&\frac{1}{N\lambda_N}\paren{\PRP{\h{N}\geq\lambda_N}-\lambda_N\ES{\frac{1}{\h{N}}\I{\h{N}\geq\lambda_N}}}
\end{eqnarray}
since channel power gains are i.i.d.  Note that $\lambda_N$ is chosen such that the average tranmit power constraint is satisfied with equality, \emph{i.e.,} $\ES{\paren{\frac{1}{\lambda_N}-\frac{1}{h_i}}^+\I{h_i=\h{N}}}=P_{\rm ave}$.  Thus, \eqref{Eq: Pave} implies that $\lambda_N \leq \frac{1}{N P_{\rm ave}}$, and $\lambda_N$ converges to zero as $N$ becomes large.  This convergence result further implies that $\lim_{N\ra\infty}\PR{\h{N}\geq\lambda_N}=1$.  Hence, it is enough to show that $\ES{\frac{1}{\h{N}}\I{\h{N}\geq\lambda_N}}$ converges to zero as $N$ tends to infinity to complete the proof.  From \eqref{Eq: hstar-concentration}, we directly have $\frac{1}{\h{N}}\I{\h{N}\geq\lambda_N}$ converging to zero in probability.  Using arguments similar to the ones used in Appendices \ref{App: TPIL} and \ref{App: IL} above, it can also be shown that $\left\{\frac{1}{\h{N}}\I{\h{N}\geq\lambda_N}\right\}_{N=1}^\infty$ is a uniformly integrable collection of random variables, which finishes the proof. 
 \end{IEEEproof} 

In the next lemma, we establish the asymptotic scaling behavior of $\ES{g_{\rm min}(N)}$, where $g_{\rm min}(N) = \min_{1\leq i\leq N}g_i$, as $N$ grows large.  Later, this result will be helpful in the process of obtaining lower bounds for the sum-rate of secondary networks under $\CoSIPILFull$ and $\CoSIPILRed$.
  \begin{lemma}\label{Lemma: gmin Scaling}
 Let $g_{\rm min}(N)=\min_{1\leq i\leq N}g_i$.  Then, 
 \begin{eqnarray}
 \lim_{N\ra\infty}\frac{\ES{g_{\min}(N)}}{F_g^{-1}\paren{\frac{1}{N}}}=\Gamma\paren{1+\frac{1}{\gamma_g}},\nonumber
 \end{eqnarray}
 where $\Gamma\paren{\cdot}$ is the Gamma function.
 \end{lemma}
 \begin{IEEEproof}
 Observe that $F_g^{-1}(0)=0$ and $\lim_{x\downarrow0}\frac{F_g\paren{\epsilon x}}{F_g\paren{x}}=\epsilon^{\gamma_g}$ for $\epsilon>0$, {\em i.e.,} see Definition \ref{Def1}.  Hence, the sequence of random variables $\left\{\frac{g_{\min}(N)}{F_g^{-1}\paren{\frac{1}{N}}}\right\}_{N=1}^\infty$ converges in distribution to a Weibull distributed random variable with shape parameter $\gamma_g$ and scale parameter 1, {\em i.e.}, to the CDF $F\paren{x}=1-\e{-x^{\gamma_g}}$ for $x\geq0$ \cite{Arnold}.  Applying Fatou's lemma, we have $\liminf_{N\ra\infty}\frac{\ES{g_{\min}(N)}}{F_g^{-1}\paren{\frac{1}{N}}}\geq \ES{W}=\Gamma\paren{1+\frac{1}{\gamma_g}}$. 
 
To show the other direction, we write $\ES{g_{\min}(N)}$ as 
\begin{eqnarray}\label{Eq: gmin-expansion}
\ES{g_{\min}(N)} &=&\int_{0}^{\epsilon_1}\paren{1-F_g(x)}^Ndx+\int_{\epsilon_1}^N\paren{1-F_g(x)}^Ndx+\sum_{i=1}^\infty\int_{N^i}^{N^{(i+1)}}\paren{1-F_g(x)}^Ndx
  \end{eqnarray}
 for all $\epsilon_1>0$. For all $\epsilon_2$ belonging to $\paren{0,1}$ and $\epsilon_1$ small enough, $\int_{0}^{\epsilon_1}\paren{1-F_g(x)}^Ndx$ can be upper bounded as 
 \begin{eqnarray}\label{Eq: Pre-Watson}
 \int_{0}^{\epsilon_1}\paren{1-F_g(x)}^Ndx&\stackrel{\rm (a)}{\leq}& \int_{0}^{\epsilon_1}\paren{1-\paren{1-\epsilon_2}\eta_g x^{\gamma_g}}^Ndx\nonumber\\
 &=& \int_{0}^{\epsilon_1}\e{N\logp{1-\paren{1-\epsilon_2}\eta x^{\gamma_g}}}dx\nonumber\\
&\stackrel{\rm (b)}{\leq}& \int_{0}^{\epsilon_1}\e{-N\paren{1-\epsilon_2}\eta_g x^{\gamma_g}}dx\nonumber\\
&=& \frac{1}{\gamma_g}\int_{0}^{\epsilon_1^{\gamma_g}}x^{\frac{1}{\gamma_g}-1}\e{-N\paren{1-\epsilon_2}\eta_g x }dx,\nonumber
\end{eqnarray}
where (a) follows from Definition \ref{Def1}, and (b) follows from that fact that $\logp{1-x} \leq -x$ for $0 \leq x < 1$. Applying Watson's lemma \cite{Murray}, we have $\lim_{N\ra\infty}\frac{1}{\gamma_g}\frac{\int_{0}^{\epsilon_1^{\gamma_g}}x^{\frac{1}{\gamma_g}-1}\e{-N\paren{1-\epsilon_2}\eta_g x}dx}{F_g^{-1}\paren{\frac{1}{N}}}=\frac{\Gamma\paren{1+\frac{1}{\gamma_g}}}{1-\epsilon_2}$ for all $\epsilon_2>0$, which implies $\limsup_{N\ra\infty}\frac{\int_{0}^{\epsilon_1}\paren{1-F_g(x)}^Ndx}{F_g^{-1}\paren{\frac{1}{N}}}\leq\Gamma\paren{1+\frac{1}{\gamma_g}}$.  To complete the proof, it is enough to show that the second and third terms in \eqref{Eq: gmin-expansion} decay to zero faster than $F_g^{-1}\paren{\frac{1}{N}}$, \emph{i.e.,} they can be asymptotically expressed as $\LO{F_g^{-1}\paren{\frac{1}{N}}}$.  $\int_{\epsilon_1}^N\paren{1-F_g(x)}^Ndx$ can be upper bounded as
 \begin{eqnarray}
 \int_{\epsilon_1}^N\paren{1-F_g(x)}^Ndx&\leq& N\paren{1-F_g(\epsilon_1)}^N\nonumber\\
 &=&N\e{-\BT{N}}. 
 \end{eqnarray}
Thus, by recalling that $F_g$ varies regularly around the origin, $\int_{\epsilon_1}^N\paren{1-F_g(x)}^Ndx=\LO{F_g^{-1}\paren{\frac{1}{N}}}$.  Similarly, for all $\epsilon_3>0$ and $N$ large enough, $\int_{N^i}^{N^{(i+1)}}\paren{1-F_g(x)}^Ndx$ can be upper bounded as
 \begin{eqnarray}
 \int_{N^i}^{N^{i+1}}\paren{1-F_g(x)}^Ndx&\leq & N^{i+1}\paren{1-F_g\paren{N^i}}^N\nonumber\\
 &\stackrel{\rm (a)}{\leq}&N^{i+1}\e{N\logp{\paren{1+\epsilon_3}\alpha_h N^{il_h}\e{-\beta_h N^{in_h}+H\paren{N^i}}}}\nonumber\\
&=&\e{-N} N^{i+1}\e{-\BT{ N^{in_h}}},\nonumber
 \end{eqnarray} 
where (a) follows from Definition \ref{Def1}. Therefore, $\sum_{i=1}^\infty\int_{N^i}^{N^{(i+1)}}\paren{1-F_g(x)}^N=\LO{F_g^{-1}\paren{\frac{1}{N}}}$, which completes the proof.
 \end{IEEEproof}
 
\subsection{Proof of Throughput Scaling in $\CoSIPILFull$}
We are now ready to establish the sum-rate scaling behavior for secondary networks under $\CoSIPILFull$.  By removing the individual power constraints, we obtain a secondary network under $\CoSILFull$.  Thus, the sum-rate of a secondary network under $\CoSIPILFull$ is upper bounded by the sum-rate of the same network under $\CoSILRed$, \emph{i.e,} $\RateIPILFull \leq \RateILFull$.  Hence, we have $\limsup_{N\ra\infty}\frac{\RateIPILFull}{\logp{N}}\leq \frac{1}{\gamma_g}$.  Similarly, by removing the average interference power constraint, we obtain a primary MAC network with individual power constraints.  Therefore, the sum-rate under $\CoSIPILFull$ is upper bounded by the sum-rate of a primary MAC network with the same individual power constraints, \emph{i.e.}, $\RateIPILFull\leq\RateIPM$. Thus, we also have $\limsup_{N\ra\infty}\frac{\RateIPILFull}{\logp{N}}\leq 1$, implying $\limsup_{N\ra\infty}\frac{\RateIPILFull}{\logp{N}}\leq\min\paren{1, \frac{1}{\gamma_g}}$.
 
To show the other direction, consider a sub-optimum power allocation policy $\SPAP{IPIL}{N}$ in which the transmit power of SU-$i$ is given by the formula $\SPC{IPIL}{N}=\epsilon N^{\min\paren{1,\frac{1}{\gamma_g}}}\I{g_i=g_{\rm min}(N)}$ for some $\epsilon>0$. We will first show that $\SPAP{IPIL}{N}$ is a feasible power allocation policy for $N$ large enough and an appropriate choice of $\epsilon$.  The average transmit power of SU-$i$ under $\SPAP{IPIL}{N}$ is equal to $\ES{\SPC{IPIL}{N}}=\epsilon N^{\min\paren{1,\frac{1}{\gamma_g}}-1}$ since the interference channel gains are i.i.d.  Thus, for $N$ large enough and a proper choice of $\epsilon$, $\SPAP{IPIL}{N}$  satisfies individual average power constraints.  The average interference power at the PBS under $\SPAP{IPIL}{N}$ is given by $\ES{\sum_{i=1}^Ng_i\SPC{IPIL}{N}}=\epsilon N^{\min\paren{1,\frac{1}{\gamma_g}}}\ES{g_{\rm min}(N)}$.  Using Lemma \ref{Lemma: gmin Scaling}, the average interference power at the PBS can be asymptotically expressed as $\ES{\sum_{i=1}^Ng_i\SPC{IPIL}{N}}=\epsilon \BO{N^{\min\paren{1,\frac{1}{\gamma_g}}-\frac{1}{\gamma_g}}}$.  Thus, for $N$ large enough and a proper choice of $\epsilon$, $\SPAP{IPIL}{N}$ meets the average interference power constraint as well, and it is a feasible power allocation policy for a secondary network under $\CoSIPILFull$.

 Let $\SRate{IPIL}{N}$ be the secondary network sum-rate under $\SPAP{IPIL}{N}$. Note that $\SRate{IPIL}{N}$ serves as a lower bound for $\RateIPILFull$ for an appropriate choice of $\epsilon$ and $N$ large enough, \emph{i.e.,} $\SRate{IPIL}{N} \leq \RateIPILFull$ for $N$ large enough.  $\hat{R}_{\rm IPIL}\paren{N}$ can be further lower bounded as
\begin{eqnarray}
\SRate{IPIL}{N}
&=&\sum_{i=1}^N\ES{\logp{1+h_i\epsilon N^{\min\paren{1,\frac{1}{\gamma_g}}}}\I{g_i=g_{\rm min}(N)}}\nonumber\\
&\geq&\min\paren{1,\frac{1}{\gamma_g}}\logp{N} + \logp{\epsilon} + \ES{\logp{h}}, \nonumber
\end{eqnarray}
where $h$ is a generic random variable with CDF $F_h(x)$.  Therefore, $\liminf_{N\ra\infty}\frac{\RateIPILFull}{\logp{N}}\geq\min\paren{1,\frac{1}{\gamma_g}}$, which completes the proof.
\subsection{Proof of Throughput Scaling in $\CoSIPILRed$} 
Since more information is available at the SBS to perform power-control and user scheduling in $\CoSIPILFull$, we have $\RateIPILRed\leq \RateIPILFull$, and  therefore $\limsup_{N\ra\infty}\frac{\RateIPILRed}{\logp{N}}\leq \min\paren{1,\frac{1}{\gamma_g}}$. To prove the other direction, consider a sub-optimum power allocation policy $\SPAP{IPIL}{\K}$ in which the transmit power of SU-$i$ is given by $\SPC{IPIL}{\K}=\epsilon N^{\min\paren{1,\frac{1}{\gamma_g}}}\I{g_i = \min_{1 \leq j \leq \K}g_{\pi(j)}}$ for some $\epsilon>0$.  Let $\SRate{IPIL}{\K}$ be the secondary network sum-rate under $\SPAP{IPIL}{\K}$.  Similar to our feasibility proof for $\SPAP{IPIL}{N}$ in $\CoSIPILFull$, we can show that for $N$ large enough and a proper choice of $\epsilon$, $\SPAP{IPIL}{\K}$ becomes a feasible power allocation policy.  Hence, $\SRate{IPIL}{\K} \leq \RateIPILRed$.  It also follows that $\liminf_{N\ra\infty}\frac{\SRate{IPIL}{\K}}{\logp{N}}\geq \min\paren{1,\frac{1}{\gamma_g}}$.  Therefore, $\liminf_{N\ra\infty}\frac{\RateIPILRed}{\logp{N}}\geq \min\paren{1,\frac{1}{\gamma_g}}$, which completes the proof.
\bibliographystyle{IEEEtran}

\end{document}